\documentclass[aps,pra,superscriptaddress,twocolumn]{revtex4}
\usepackage{graphicx,amsmath,amssymb,amsfonts,latexsym,color,dcolumn,bm}

\newcommand{\beq}{\begin{equation}}
\newcommand{\eeq}{\end{equation}}
\newcommand{\bea}{\begin{eqnarray}}
\newcommand{\eea}{\end{eqnarray}}
\providecommand{\abs}[1]{\left\lvert#1\right\rvert}

\providecommand{\bra}[1]{\langle #1 \rvert}
\providecommand{\ket}[1]{\lvert #1 \rangle}

\newcommand{\bcomment}[1]{\textcolor{blue}{#1}}
\newcommand{\news}[1]{\textcolor{red}{ #1}}
\usepackage{amsfonts,amssymb}
\usepackage{dsfont}
\usepackage{physics}
\usepackage{tocvsec2}

\usepackage{appendix}

\usepackage{tikz}

\usepackage{amsmath}
\usepackage{bbold}
\usepackage{hyperref}
\hypersetup{
     colorlinks=true,      
    linkcolor=blue,        
    citecolor=blue,        
    filecolor=magenta, 
    urlcolor=black          
}

\usepackage{comment}

\begin{document}

\author{N. Fabre\footnote{n.fabre@cent.uw.edu.pl}}
\affiliation{Centre for Quantum Optical Technologies, Centre of New Technologies, University of Warsaw, ul.  Banacha 2c, 02-097 Warszawa, Poland
}

\author{S. Felicetti}
\affiliation{Istituto di Fotonica e Nanotecnologie, Consiglio Nazionale delle Ricerche (CNR-IFN), Via Cineto Romano 42, I-00156 Rome, Italy}

\date{\today}
\begin{abstract}

Hong-Ou-Mandel interferometry takes advantage of the quantum nature of two-photon interference to increase the resolution of precision measurements of time-delays. Relying on few-photon probe states, this approach is applicable also in cases of extremely sensible samples and it achieves attosecond (nanometer path length) scale resolution, which is relevant to cell biology and two-dimensional materials. Here, we theoretically analyze how the precision of Hong-Ou-Mandel interferometers can be significantly improved by engineering the spectral distribution of two-photon probe states. In particular, we assess the metrological power of different classes of biphoton states with non-Gaussian time-frequency spectral distributions, considering the estimation of both time- and frequency-shifts. We find that grid states, characterized by a periodic structure of peaks in the chronocyclic Wigner function, can outperform standard biphoton states in sensing applications. The considered states can be feasibly produced with atomic photon sources, bulk non-linear crystals and integrated photonic waveguide devices.
 \end{abstract}
\vskip2pc 

\title{Parameter estimation of time and frequency shifts \\with generalized HOM interferometry}

\maketitle

\section{Introduction}
Quantum technologies can reach an intrinsic advantage over classical devices in sensing applications, by exploiting nonclassical features of photonic, atomic and solid-state systems. A paradigmatic experiment unveiling nonclassical features of light is given by Hong, Ou and Mandel  (HOM) interferometry \cite{HOM}, which consists in the interference of two identical particles onto a beam-splitter, followed by a coincidence measurement. HOM interference has been exploited in a variety of applications in quantum optics~\cite{ou2007multi,aspuru2012photonic,gianani2020measuring}. The characteristic dip in the coincidence rate is a direct consequence of the particle indistinguishability, and it has been exploited to assess the purity of single photons~ \cite{Cassemiro_2010,Mosley2008} and to characterize solid-state photon sources~\cite{Kim2016,aharonovich2016}. 
Multi-dimensional HOM interference has also been demonstrated making use of spatial-temporal~\cite{Devaux2020} and spectral-temporal~\cite{Gerrits2015,Jin2015,Gianani2018,Thiel2020,chen2020temporal,zhang2021high} modes.
Of particular interest for the present work, HOM interferometry makes it also possible to engineer the joint spectrum of photon pairs. For example, fermionic or bosonic statistics can be simulated generating entangled biphoton states~\cite{sansoni_two-particle_2012,francesconi_engineering_2020}. In the generalized HOM scheme an additional frequency shift is applied on one of the two arms of the interferometer, while the usual phase shift is applied on the other one. Generalized HOM interferometry has been used to characterize the chronocyclic Wigner function of photon pairs generated via parametric processes~\cite{douce_direct_2013,boucher_toolbox_2015}. This paved the way to the generation and manipulation of entangled  grid states defined in the time-frequency space~\cite{PhysRevA.102.012607,PhysRevLett.91.163602}. Complex engineering in two-dimensional frequency space has also been demonstrated with bulk~\cite{walborn_multimode_2003, jin_quantum_2018} and integrated devices~\cite{francesconi_engineering_2020}.

HOM interferometry is a compelling approach for sensing applications, as it is robust to changes of the relative phase between the photons, and so it does not require demanding stabilization stages that are usually needed in classical interferometers.  In this context, the HOM interference has been used 
to measure polarization mode dispersion~\cite{Branning2000}, to perform quantum optical coherence tomography~\cite{Abou2002,Nasr2003,nasr2004dispersion,lopez2012}, and in nonlinear spectroscopy~\cite{mukamel2020roadmap,Dorfman2021}. These approaches are based on the detection of the minimum of the coincidence rate, and so the ultimate limit of the precision is given by the width of the HOM dip, which is set by the photon coherence time. Very recently, a new approach has been proposed applying the framework of quantum metrology in HOM interferometry. This approach allows one to tune the path delay to the value that minimizes the estimation error, and it made it possible to reach attosecond resolution in precision measurements of time delays~\cite{lyons_attosecond-resolution_2018,chen_hong-ou-mandel_2019}. Such an interferometer was also used for polarization measurement \cite{harnchaiwat_tracking_2020}, and the effect of the time resolution of the photodetectors was investigated in \cite{scott_beyond_2020}, paving the way to calibration-free protocols. Furthermore,  it has been demonstrated~\cite{chen_hong-ou-mandel_2019} that using a frequency-entangled photon pair can result in a significant increase of the estimation precision.   

In this article, we show how the precision of quantum parameter-estimation protocols based on generalized HOM interferometry can be improved by engineering the the spectral distribution of photon pairs. We analyze the estimation of time and frequency displacements considering biphoton states that can be produced with different platforms, such as trapped atomic system, bulk non-linear crystal and optical integrated waveguide devices. 
We derive analytical expressions for the quantum Fisher information (QFI), in order to assess the metrological power of different kinds of non-Gaussian continuous-variable states defined in the time-frequency space. Then, to analyze the estimation precision achievable with HOM interferometers, we derive the Fisher information (FI) relative to the measurement of the coincidence number.
In particular, we compare the performances achievable using different classes of biphoton states, considering (1) states with a Gaussian chronocyclic Wigner distribution, (2) non-Gaussian coherent superpositions of two Gaussian states, namely time-frequency cat-like states~\cite{PhysRevA.102.023710}, and (3) grid states with a chronocyclic Wigner function characterized by a periodic structure in both time and frequency axis. We also show how the addition of a frequency or temporal chirp can improve the metrological power for multiparameter estimation. Notably, we find that grid states can outperform simpler biphoton states in both frequency- and temporal-displacement estimation. 
Notice that Grid states represent a time-frequency analogue of Gottesman, Kitaev and Preskill (GKP) states defined in the standard quadrature-momentum phase space~\cite{gottesman_encoding_2001}, which also represent a valuable resource for quantum sensing~\cite{PhysRevA.95.012305,LEGERO2006253}. Thus, our analysis links the framework of HOM interferometry and the field of quantum sensing with non-Gaussian continuous-variable states~\cite{PhysRevA.97.032116,PhysRevLett.122.090503}.


Let us now briefly comment on the quantum nature of the considered protocols. The common approach to assess  quantum advantage in parameter estimation protocols consists in analyzing the scaling of the achievable precision as a function of the number $N$ of probes~\cite{QMetrologyGiovannetti,QMetrologyToth}. The estimation precision is limited by the Cramer-Rao bound, that is, the variance of any unbiased estimator is bounded by the reciprocal of the FI. When $N$ uncorrelated states are used as a probe, the FI grows at most as $~\sqrt{N}$, and so the optimal scaling of the estimation error achievable with a classical protocol is given by $~1/\sqrt{N}$. This fundamental limit can be overcome using quantum resources, such as squeezing~\cite{Kwon2019,Garbe2019} or quantum correlations between the probe systems~\cite{QMetAtomicPezze,Hyllus2010}, which make it possible to reach a $~1/N$ scaling of the estimation error, known as Heisenberg limit. However, in quantum estimation protocols based on HOM interferometry~\cite{lyons_attosecond-resolution_2018,chen_hong-ou-mandel_2019,scott_beyond_2020,harnchaiwat_tracking_2020} the input is composed of two-photon states, and so for each run of the protocol the number of probes is fixed to $N=2$. In this case, quantum metrological advantage is not coming from the number of entangled photons, but it resides in the single-photon nature of the considered states. Indeed, the HOM dip is limited to one half for classical fields while it can reach zero for biphoton states, and so the visibility of the interference signal is intrinsically higher when quantum resources are available. This approach makes it in principle possible to achieve relevant quantum advantage also when only a very limited number of photons can be used. A compelling example where the intensity of the probe field can be relevant is biological sensing~\cite{TAYLOR20161}, where a low photon flux must be used to avoid high-power fields that can damage the sample.  The proposed quantum-sensing scheme can achieve frequency resolutions comparable to the typical size of lipids $7.5$~nm and water molecules $0.27$~nm.

The paper is organized as follows. In Sec.~\ref{sectiontwo}, we introduce the formalism and the theoretical tools used in the rest of the manuscript. First, we describe the generalized HOM experiment using the chronocyclic Wigner function formalism; then, we introduce the FI and the QFI, and we derive general expressions valid for HOM interferometry with biphoton states. Finally, we analyze the metrological power of states with a Gaussian distribution in the time-frequency space. In Sec.~\ref{sectionthree}, we evaluate the metrological power of frequency and time cat-like biphoton states. We first derive the QFI, and then the FI for generalized HOM measurement, and we analyze how the multi-paramementer estimation precision can be improved adding a frequency or temporal chirp. In Sec.~\ref{sectionfour}, we derive the expression of the QFI of different time-frequency grid states and show that it can reach higher values compared to  time-frequency Gaussian and cat-like states. Also in this case, we analyze the FI for HOM measurement and the effect of frequency or temporal chirp. Finally, in Sec.~\ref{conclusions} we provide a brief summary of our results and comments on future perspectives.

\section{Quantum metrology with time-frequency biphoton states}\label{sectiontwo}

Let us review the description of the generalized HOM experiment \cite{douce_direct_2013} where, in addition to the usual time displacement  applied  in one of the arms of the interferometer, a frequency shift is applied on the other arm, before the recombination into the beam-splitter. We then introduce the Fisher information (FI) and the Quantum Fisher Information (QFI) and proceed to the evaluation of the QFI for a simple time-frequency Gaussian state.

\subsection{Generalized HOM experiment and chronocyclic Wigner function formalism}

We consider a two-photon pair with orthogonal polarization, called  signal and idler photons, produced by a type-II SPDC process, by a three or four wave mixing non-linear effect. Since the photons have opposite polarizations, they can be separated deterministically with a polarizing beam-splitter into two spatial paths labeled $a$ and $b$.  After the PBS, the wavefunction of this  pure quantum state can be written as,
\begin{equation}
\ket{\psi}= \iint d\omega_{s}d\omega_{i}\text{JSA}(\omega_{s},\omega_{i}) \hat{a}^{\dagger}(\omega_{s})\hat{b}^{\dagger}(\omega_{i}) \ket{0},
\end{equation}
where $\ket{0}$ is the vacuum state, while $\hat{a}^{\dagger},\hat{b}^{\dagger}$ are the creation operator of a single photon in the spatial mode $a$ and $b$, respectively. We denote with $\omega_{s/i}$  the frequency of the signal and idler photons. The joint spectral amplitude (JSA) can be written as,
\begin{equation}\label{decompositionJSA}
\text{JSA}(\omega_{s},\omega_{i})=f_{+}(\omega_{+}-\omega_{p}) f_{-}(\omega_{-}),
\end{equation}
where $f_{+,-}(\omega_{\pm})$ is the function characterizing the energy conservation (resp. the phase matching condition), and is a function of the collective variables $\omega_{\pm}=\omega_{s}\pm \omega_{i}$.  For the sake of simplicity, the degeneracy frequency $\omega_{p}$ of the SPDC process will be set to zero with no loss of generality. The joint spectral intensity (JSI) is defined as the absolute squared value of the JSA and it corresponds to the joint probability of measuring a signal photon at frequency $\omega_{s}$ and an idler photon at frequency $\omega_{i}$. Then, with the decomposition defined by Eq.~(\ref{decompositionJSA}), the wave function can be written:
\begin{equation}\label{factorization}
\ket{\psi}=\iint  d\omega_+d\omega_- f_{+}(\omega_+) f_{-}(\omega_-)  \ket{\omega_s} \ket{\omega_i} .
\end{equation}
The range of integration is  $\mathds{R}$ because we are dealing with the collective variables $\omega_{\pm}$. The normalization of the wave function imposes that $\iint d\omega_{s} d\omega_{i} \abs{\text{JSA}(\omega_{s},\omega_{i})}^{2} =\int d\omega_{+} \abs{f_{+}(\omega_{+})}^{2} \int d\omega_{-} \abs{f_{-}(\omega_{-})}^{2}$ =1 . 
We define the joint temporal amplitude (JTA) as the Fourier transform of the JSA, $\text{JTA}(t_{s},t_{i})=\iint d\omega_{s} d\omega_{i} \text{JSA}(\omega_{s},\omega_{i}) e^{i(\omega_{s}t_{s}+\omega_{i}t_{i})}$. Similarly to the JSI, the joint temporal intensity (JTI) is defined as the absolute square value of the JTA, and it corresponds to the probability of measuring the signal photon at arrival time $t_{s}$ and the idler photon at $t_{i}$. We will denote the Fourier transform of the functions $f_{+}(\omega_{\pm})$ with a tilde as follows $\tilde{f}_{\pm}(t_{\pm})=\int_{\mathds{R}} d\omega_{\pm} e^{i\omega_{\pm}t_{\pm}} f_{\pm}(\omega_{\pm})$
 where we defined the temporal collective variables $t_{\pm}=t_{s}\pm t_{i}$. 

In general, continuous-variable systems are conveniently described in terms of the Wigner distribution.
The analogue of the Wigner function defined for time-frequency continuous variables is called chronocyclic Wigner function
 \cite{PhysRevA.102.012607,brecht_characterizing_2013}, and for biphoton states it can be written under the form,
\begin{multline}\label{Wignerbiph}
W_{\hat{\rho}}(\omega_{s},\omega_{i},t_{s},t_{i})= \frac{1}{\pi^{2}}\iint d\omega_{s}'d\omega_{i}' e^{2i\omega_{s}' t_{s}}e^{2i\omega'_{i} t_{i}}  \\
\cross \bra{\omega_{s}-\omega_{s}',\omega_{i}-\omega'_{i}} \hat{\rho} \ket{\omega_{s}+\omega_{s}',\omega_{i}+\omega'_{i}},
\end{multline}
where $\hat \rho$ is the density matrix and the expression is normalized to one, $\iint \iint d\omega_{s} dt_{s} d\omega_{i} dt_{i}  W_{\hat{\rho}}(\omega_{s},\omega_{i},t_{s},t_{i}) =1$.  
The four marginals of the chronocyclic Wigner distribution correspond to physical quantities that are experimentally accessible~ \cite{maclean_reconstructing_2019},
\begin{align}
\iint W_{\hat{\rho}}(\omega_{s},\omega_{i},t_{s},t_{i}) d\omega_{s} d\omega_{i}=\text{JTI}(t_{s},t_{i}),\\
\iint W_{\hat{\rho}}(\omega_{s},\omega_{i},t_{s},t_{i}) dt_{s} dt_{i}=\text{JSI}(\omega_{s},\omega_{i}),\\
\iint W_{\hat{\rho}}(\omega_{s},\omega_{i},t_{s},t_{i}) d\omega_{s/i} dt_{i/s}= \text{JTSI}(\omega_{s/i},t_{i/s}).
\end{align}
We defined the joint temporal-spectral intensity (JTSI) that corresponds to the probability of measuring one photon with arrival time $t$ and the other with frequency $\omega$.   Since we consider a pure state $\hat{\rho}=\ket{\psi}\bra{\psi}$, and with the factorization defined in Eq.~(\ref{factorization}), the chronocyclic Wigner distribution of the photon pair can be factorized as follows,
\begin{equation}\label{factorization2}
W_{\hat{\rho}}(\omega_{s},\omega_{i},t_{s},t_{i})=W_+(\omega_{+},t_{+})W_-(\omega_{-},t_{-}),
\end{equation}
where $W_{\pm}(\omega_\pm,t_\pm)=\frac{1}{\pi} \int e^{2i\omega't_{\pm}} f_{\pm}(\omega_{\pm}-\omega')f^{*}_{\pm}(\omega_{\pm}+\omega')d\omega'$. As discussed in the following, the generalized HOM experiment corresponds to a direct measurement  of the phase-matching function $W_{-}(\omega_{-},t_{-})$. The marginal of the chronocyclic Wigner distribution $W_{\pm}(\omega_\pm,t_\pm)$ are respectively the squared absolute value  of the spectral and temporal distribution of the energy conservation (resp. phase matching function) $\abs{f_{\pm}(\omega_{\pm})}^{2}$ and $\abs{\tilde{f}_{\pm}(t_{\pm})}^{2}$. The inverse transform from the chronocyclic Wigner distribution to the function $f_{\pm}$ can be cast in the form:
\begin{equation}
f_{\pm}(\omega_{\pm})=\frac{1}{f_{\pm}^{*}(0)} \int W_\pm (\omega_\pm/2,t)e^{i\omega_\pm t} dt.
\end{equation}
Note that each individual chronocyclic Wigner distribution $W_{\pm}$ is normalized to one. We can give some examples of the function $f_{\pm}$. The  time-frequency analogue of EPR states (\textit{i.e.} two-mode squeezed state with infinite squeezing parameter) will be defined as $\abs{f_{+}(\omega_+)}^{2}=\delta(\omega_+)$ for anticorrelated frequency photon pair, while  for correlated frequency photon pair, we will have $\abs{f_{-}(\omega_-)}^{2}=\delta(\omega_-)$. A physically-relevant approximation to these states can be obtained replacing the Dirac distribution with Gaussian functions.

\begin{figure*}
\begin{center}
 \includegraphics[scale=0.3]{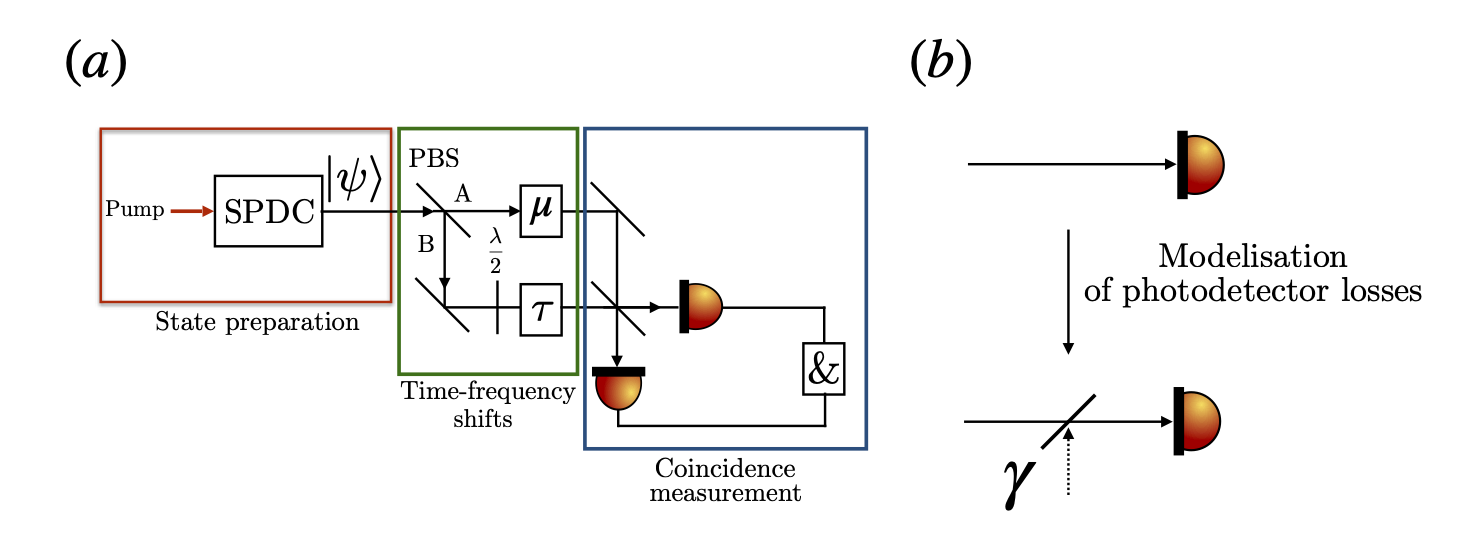}
 \caption{\label{HOMexperiment} (a) Schematic of the generalized HOM experiment. A monochromatic pump crosses a non-linear optical medium generating a photon pair by a type-II SPDC described by the wavefunction $\ket{\psi}$. The photon pair is then spatially separated by a polarizing beam-splitter. A frequency $\mu$ (resp. a time $\tau$) delay is applied in spatial port A and B. Finally, a non-time resolved coincidence measurement is performed. (b) Modelisation of the photon losses of the photo-detector  with a beam-splitter with reflectivity $\gamma$ which quantifies the photon losses.  }
 \end{center}
\end{figure*}

Let us now describe the HOM interferometer using the formalism we have introduced. After the separation of the photon pair generated by SPDC (see Fig.~\ref{HOMexperiment}), two unitary operations are performed: a frequency displacement  in path $a$ and a time displacement operation in path $b$. The wavefunction of the photon pair after these displacements operations can be written as:
\begin{multline}\label{wavefunctionfirstwritingdisplacement}
\ket{\psi(\tau,\mu)}= \iint d\omega_{s}d\omega_{i} e^{i\omega_{s}\tau} f_{+}(\omega_{+}+\mu)f_{-}(\omega_{-}+\mu)\\
 \hat{a}^{\dagger}(\omega_{s})\hat{b}^{\dagger}(\omega_{i}) \ket{0}.
\end{multline} 
After changing the polarization with a half-plate of one photon of the pair, the two photons are then recombined into a balanced beam-splitter. By selecting only the coincidence events, the post-selected wave function can be written as,
\begin{multline}
\ket{\psi(\tau,\mu)}=\frac{1}{2}[ \iint d\omega_{s}d\omega_{i} f_{+}(\omega_{+}+\mu) (e^{i\omega_{s}\tau} f_{-}(\omega_{-}+\mu)\\
 -e^{i\omega_{i}\tau}f_{-}(-\omega_{-}+\mu) )\hat{a}^{\dagger}(\omega_{s})\hat{b}^{\dagger}(\omega_{i}) \ket{0}].
\end{multline} 
Assuming a flat frequency response for the detectors, the coincidence probability is: $I(\tau,\mu)=\iint d\omega_{s} d\omega_{i} \abs{\bra{\omega_{s},\omega_{i}}\ket{\psi(\tau,\mu)}}^{2}$ (see \cite{PhysRevA.102.023710} for a mathematical proof), and so we can write
\begin{multline}
I(\mu,\tau)=\frac{1}{2}(1-\int d\omega_{+} \abs{f_{+}(\omega_{+})}^{2}\int d\omega_- e^{2i\omega_{-} \tau} \\
\cross f_{-}(\omega_{-}+\mu)f^{*}_{-}(-\omega_{-}+\mu)).
\end{multline}
Normalization of the wavefunction sets  $\int d\omega_{+} \abs{f_{+}(\omega_{+})}^{2}=1$. The coincidence rate can also be expressed in terms of the chronocyclic Wigner distribution~\cite{brecht_characterizing_2013, PhysRevA.102.012607}  $W_{-}$ of the phase matching function of the SPDC process,
\begin{equation}\label{coincidenceprobability}
I(\mu,\tau)=\frac{1}{2}(1- \pi W_{-}(\tau,\mu)).
\end{equation}
The last equation shows that the generalized HOM experiment allows one to perform the tomography of the phase-matching function. Being $I_B$ the probability of bunching, that is of two photons exiting the interferometer through the same port, we have that $I_{B}(\mu,\tau)+I(\mu,\tau)=1$. When the photons are fully distinguishable, the coincidence rate is given by 1/2, and the bunching and coincidence events happen with the same probability. Furthermore,  when $I(\mu,\tau)>1/2$ the chronocyclic Wigner distribution is negative, which is an entanglement witness \cite{Eckstein:08}.

Let us now comment on how it is possible to perform measurements of time-frequency continuous variables in practice.
To our knowledge, the full reconstruction of the chronocyclic Wigner distribution of the phase matching function $W_{-}$ has not yet been experimentally implemented. Since the spectral bandwidth in \cite{Orieux:11,francesconi_engineering_2020,ferreri_spectrally_2020} for the optical integrated systems is in the order of 50 GHz, and in bulk system \cite{chen_hong-ou-mandel_2019} is 250 GHz, the technical challenge is in the required sampling of such a distribution. The smallest  frequency  shift achievable with nowadays  electro-optic modulators is limited to 15 GHz in the telecom band \cite{chen_single-photon_2021}. Another way for implementing such frequency shift for  wider spectral distributions consists in introducing a dynamical shift of the eigenfrequencies or eigenmodes of optical resonators or waveguides \cite{kuramochi_single-photon_2016}, which corresponds to a tuning of the photon frequency. A waveguide-based optomechanical frequency shift can displace the frequency of telecom wavelength single photon up to 150 GHz, thanks to the piezoelectric effect (see Ref. \cite{fan_integrated_2016}). It has also been demonstrated that such shift does not deteriorate the HOM visibility.
The experimental possibility of implementing  frequency displacements motivates the interest in a protocol that estimates the amount of frequency shifts. HOM interferometry has already proven to be a valuable approach for temporal displacement estimation~\cite{lyons_attosecond-resolution_2018,chen_hong-ou-mandel_2019}, and so the generalized HOM setting represents a promising method to simultaneously estimate temporal  $\tau$ and frequency $\mu$ shifts.

\subsection{Classical and Quantum Fisher information}\label{QFIANDFI}

The precision of a parameter estimation protocol is quantified by the covariance  $\text{cov}(\vec{\alpha})$, where $\vec{\alpha}=(\tau,\mu)$ is a vector of the parameters to be estimated. The optimal limit of achievable precision is set by the Cramer-Rao (CR) bound $\text{cov}(\vec{\alpha}_{\text{CR}})$, if only classical resources are available. When quantum resources are available, the optimal measurement can saturate the quantum Cramer-Rao bound (QCR) $\text{cov}(\vec{\alpha}_{\text{QCR}})$. In particular we can write 
\begin{equation}\label{variancebound}
\text{cov}(\vec{\alpha}) \geq \text{cov}(\vec{\alpha}_{\text{CR}})=\frac{1}{N} (F)^{-1} \geq  \frac{1}{N} ({\cal{F}}^{\text{tot}})^{-1}=\text{cov}(\vec{\alpha}_{\text{QCR}}),
\end{equation}
where $N$ is the number of repetitions of the experiments, $F$ is the Fisher information and $\mathcal F^{\text{tot}}$ is the quantum Fisher information. The covariance is defined as $\text{cov}(\omega t)=\langle \omega t \rangle -\langle \omega\rangle \langle t \rangle$.
 Here, we consider the estimation of time and frequency displacements defined as it follows. The time (resp. frequency) displacement operation can be explicitly written~\cite{PhysRevA.102.012607} as ${\cal{\hat{D}}}(\tau)=\int \hat{a}^{\dagger}(t+\tau)\hat{a}(t) dt$, $\hat{D}(\mu)=\int d\omega \hat{a}^{\dagger}(\omega+\mu) \hat{a}(\omega)$. 
After the displacement operations, the wave function is described by Eq.~(\ref{wavefunctionfirstwritingdisplacement}). For such a pure state $\ket{\psi(\mu,\tau)}\bra{\psi(\mu,\tau)}$ , the Quantum Fisher information can be cast under the form \cite{liu_quantum_2020}:
\begin{equation}
\label{QFIdef}
{\cal{F}}^{\text{tot}}=\begin{pmatrix}
{\cal{F}}^{\text{tot}}_{\tau\tau} & {\cal{F}}^{\text{tot}}_{\mu\tau} \\
{\cal{F}}^{\text{tot}}_{\mu\tau} & {\cal{F}}^{\text{tot}}_{\mu\mu} 
\end{pmatrix}
\end{equation}
where the coefficients are
\begin{multline}\label{QFIpurestate}
{\cal{F}}^{\text{tot}}_{\mu\tau}=4(\bra{\frac{\partial \psi(\tau)}{\partial \tau}}\ket{\frac{\partial \psi(\mu)}{\partial \mu}}\\
-\abs{\bra{\psi(\mu)}\ket{\frac{\partial \psi(\mu)}{\partial \mu}}\bra{\psi(\tau)}\ket{\frac{\partial \psi(\tau)}{\partial \tau}}}).
\end{multline}
The QFI matrix elements can be expressed in two ways. The first is:
\begin{align}
{\cal{F}}^{\text{tot}}_{\tau\tau}=& 4\text{Var}(\omega_{i})\\
{\cal{F}}^{\text{tot}}_{\mu\mu}=& 4\text{Var}(t_{s})\\
{\cal{F}}^{\text{tot}}_{\mu\tau}=& 4 \text{cov}(\omega_{i}t_{s})
\end{align}
where for instance the variance is $\text{Var}(\omega_{i})=\langle \omega^{2}_{i} \rangle - \langle \omega_{i} \rangle^{2}$  defined with a statistical average, taken with respect to the chronocyclic Wigner distribution of the photon pair (see Eq.~(\ref{Wignerbiph})). The second expression of the QFI is relevant when we consider the factorization of the Wigner distribution as in Eq.~(\ref{factorization2}):
\begin{align}
{\cal{F}}^{\text{tot}}_{\tau\tau}=&4(\text{Var}(\omega_{+})+\text{Var}(\omega_{-}))=4({\cal{F}}^{+}_{\tau\tau}+{\cal{F}}^{-}_{\tau\tau})\\
{\cal{F}}^{\text{tot}}_{\mu\mu}=&4(\text{Var}(t_{+})+\text{Var}(t_{-}))=4({\cal{F}}^{+}_{\mu\mu}+{\cal{F}}^{-}_{\mu\mu})\\
{\cal{F}}^{\text{tot}}_{\mu\tau}=& - 4\text{cov}(\omega_{-}t_{-})={\cal{F}}^{-}_{\mu\tau}
\end{align}
the first two equations will be called the temporal and the frequency QFI (see Appendix \ref{fulldemonstrationQFI} for more details). The QFI part ${\cal{F}}_{-}$ which depends only on the phase-matching function will be now written ${\cal{F}}$. Here the variances denote a statistical average taken over the chronocyclic Wigner distribution of the phase matching function:
\begin{align}\label{variancephasematc}
\langle \omega_{\pm}^{\alpha} t_{\pm}^{\beta} \rangle = \iint d\omega_{\pm} dt_{\pm} \omega_{\pm}^{\alpha} t_{\pm}^{\beta} W_{\pm}(\omega_{\pm},t_{\pm}).
\end{align}
 Note that ${\cal{F}}$ does not depend on the parameters to estimate $\tau$ or $\mu$, but only on the parameters of the input wave function. Besides, the knowledge of one variance affects the knowledge of the other, which is a consequence of the non-commutativity of time and frequency displacement operators~\cite{PhysRevA.102.012607}. The origin of the non-commutativity of these operators are the commutation relations of creation and annihilation operators, and so it is a feature of few-photon states and it has no counterpart for classical fields. The Cramer-Rao bound is written in terms of the inverse of the QFI matrix,
\begin{align}
\tilde{{\cal{F}}}_{\tau\tau}=\frac{{\cal{F}}_{\mu\mu}}{{\cal{F}}_{\tau\tau}{\cal{F}}_{\mu\mu}-{\cal{F}}_{\mu\tau}^{2}}\\
\tilde{{\cal{F}}}_{\mu\mu}=\frac{{\cal{F}}_{\tau\tau}}{{\cal{F}}_{\tau\tau}{\cal{F}}_{\mu\mu}-{\cal{F}}_{\mu\tau}^{2}}\\
\tilde{{\cal{F}}}_{\mu\tau}=\frac{-{\cal{F}}_{\mu\tau}}{{\cal{F}}_{\tau\tau}{\cal{F}}_{\mu\mu}-{\cal{F}}_{\mu\tau}^{2}}.
\end{align}
 When ${\cal{F}}_{\mu\tau}=0$, the inverse of the QFI is simply a diagonal matrix of the inverse of ${\cal{F}}_{\mu\mu}$ and ${\cal{F}}_{\tau\tau}$.\\

We then consider the Fisher information \cite{liu_quantum_2020}, which is defined for a fixed measurement basis, and which can be evaluated as
 \begin{equation}\label{generalexpressionfisher}
 F_{\mu \tau}(\alpha)=\sum_{y} \frac{\partial_{\mu} P(y| \alpha)\partial_{\tau} P(y|\alpha)}{P(y,\alpha)},
 \end{equation}
 where $P$ is the probability of getting the measurement result $y$ given the parameter $\alpha$ verifying the  condition $\sum_{\alpha}P(y|\alpha)=1$. For the generalized HOM experiment, the sum in Eq.~(\ref{generalexpressionfisher}) is taken with respect to the zero, bunching and coincidence events \cite{lyons_attosecond-resolution_2018,chen_hong-ou-mandel_2019}:
\begin{multline}\label{generalexpressionFisherinfo}
F_{\mu\tau}(\tau,\mu)=\frac{\partial_{\tau} P_{2}(\tau,\mu) \partial_{\mu} P_{2}(\tau,\mu)}{P_{2}(\tau,\mu)}+\frac{\partial_{\tau} P_{1}(\tau,\mu) \partial_{\mu} P_{1}(\tau,\mu)}{P_{1}(\tau,\mu)}\\+\frac{\partial_{\tau} P_{0}(\tau,\mu) \partial_{\mu} P_{0}(\tau,\mu)}{P_{0}(\tau,\mu)},
\end{multline}
where $P_{2}$ is the coincidence rate,  $P_1$ the single count probability, and $P_0$ the no-count probability.  The finite efficiency of photodetectors  is modelized introducing  two beam-splitters of reflectivity $\gamma$ before the photo-detectors \cite{lyons_attosecond-resolution_2018,chen_hong-ou-mandel_2019} (see Fig.~\ref{HOMexperiment}):
\begin{equation}\label{linkproba}
\begin{pmatrix}
P_{0}(\tau,\mu) \\ P_{1}(\tau,\mu) \\ P_{2}(\tau,\mu)
\end{pmatrix}
=
\begin{pmatrix}
\gamma^{2} & \gamma^{2} \\
2\gamma(1-\gamma) & 1-\gamma^{2}\\
1-2\gamma(1-\gamma)-\gamma^{2} & 0
\end{pmatrix}
.
\begin{pmatrix}
I(\tau,\mu)\\
I_{B}(\tau,\mu)
\end{pmatrix}.
\end{equation}
Therefore, the probabilities can be expressed as:
\begin{align}\label{listprobability}
P_{0}(\tau,\mu)=& \ \gamma^{2},\\
P_{1}(\tau,\mu)=&\frac{1}{2}(1-\gamma)^{2}(\frac{1+3\gamma}{1-\gamma}+\pi W_{-}(\tau,\mu)),\\
P_{2}(\tau,\mu)=&\frac{1}{2}(1-\gamma)^{2}(1-\pi W_{-}(\tau,\mu)).
\end{align}
Then, we can evaluate the Fisher information, defined by Eq.~(\ref{generalexpressionFisherinfo}):
\begin{multline}
F_{\mu\tau}(\mu,\tau)=\frac{\pi^{2}}{2}(1-\gamma)^{2} \partial_{\tau}W_-(\mu,\tau)\partial_{\mu}W_-(\mu,\tau)\\
\cross (\frac{1}{1-\pi W_{-}(\mu,\tau)}+\frac{1}{\frac{1+3\gamma}{1-\gamma}+\pi W_-(\mu,\tau)}).
\end{multline}

 For single parameter estimation, we are interested in the diagonal element of the Fisher information matrix.  For time-displacement estimation, when the frequency shift is set to zero, the FI reads as:
\begin{multline}\label{fisherinformationdeveloped}
F_{\tau\tau}(0,\tau) \equiv  F_{\tau\tau}(\tau) =\frac{\pi^{2}}{2}(1-\gamma)^{2} (\partial_{\tau}W_-(0,\tau))^{2}\\
\cross (\frac{1}{1-\pi W_{-}(0,\tau)}+\frac{1}{\frac{1+3\gamma}{1-\gamma}+\pi W_-(0,\tau)}).
\end{multline}
In the limit of zero losses $\gamma\rightarrow 0$ and when the parameters $\tau,\mu\rightarrow 0$, the FI $F_{\tau\tau}(\tau)$ reaches the variance ${\cal{F}}^{-}_{\tau\tau}=\text{Var}(\omega_{-})$ for an even phase matching function, which means that the measurement is optimal (see Appendix \ref{generalcasefisher}). In particular, it is the case for a Gaussian phase matching function \cite{lyons_attosecond-resolution_2018}, or for the frequency cat-like state \cite{chen_hong-ou-mandel_2019}, which is discussed in more details in Sec.\ref{QFIsinglecolorsection} and Sec.~\ref{sectionthree}. 
A measurement strategy that saturates the Cramer-Rao bound $\delta\tau= \delta \tau_{\text{CR}}$ in Eq.~(\ref{variancebound}) has been introduced in \cite{lyons_attosecond-resolution_2018}, here we extend it to the multi-parameter case in Appendix \ref{SaturationCramerRao}. Notice that, by taking into account the losses only at the detection stage, we have neglected the photon losses between the generation of the pair and the beam-splitter, which would have as consequence to create a mixed state in the particle-number degree of freedom.  Also,  in such a case, the QFI would be no longer the variances of the chronocyclic Wigner distribution. 
Notice that one could increase the total QFI by increasing $\text{Var}(\omega_{+})$ with the same strategies that we are going to develop in this paper for increasing the term $\text{Var}(\omega_{-})$. However, we will be interested in this paper only in the $f_{-}$ part of the biphoton wavefunction because the information of $f_+$ is not accessible with generalized HOM interferometry. Accordingly, here we will consider only the QFI corresponding to the covariances of the anti-correlated frequency and time variables.

\subsection{Quantum Fisher information of a time-frequency Gaussian state}\label{QFIsinglecolorsection}
In order to evaluate the performance of various time-frequency entangled photon pair, we will use as benchmark the precision achievable with a time-frequency Gaussian state, that is a state with a Gaussian chronocyclic Wigner distribution $W_{-}$.
The wavefunction of a time-frequency Gaussian state can be written as $\ket{\psi}=\iint d\omega_{s}d\omega_{i}f_{+}(\omega_{+}) f^{\sigma}_{-}(\omega_{-}-\Delta) \ket{\omega_s}\ket{\omega_i}$. The part of the spectrum depending on $f_{+}$ is not exploitable within our metrological protocol since the generalized HOM interferometer does not give any information about  this function. The function $f^{\sigma}_{-}$ is a Gaussian function of width $\sigma$ and centered at $\Delta=\omega_{1}-\omega_{2}$. The chronocyclic Wigner distribution of the associated phase-matching function  is also Gaussian and can be written as:
\begin{equation}\label{simpleGaussian}
W_-(\omega,t)= \text{exp}(-\frac{(\omega-\Delta)^{2}}{\sigma^{2}})\text{exp}(-t^{2}\sigma^{2}).
\end{equation}
The quantum Fisher information matrix as defined in Eq.~\eqref{QFIdef} is given by,
\begin{equation}\label{onecolorfisher}
{\cal{F}}=\frac{1}{2}\begin{pmatrix}
\sigma^{2} & 0 \\ 0 & 1/\sigma^{2} 
\end{pmatrix}.
\end{equation}
Then, the higher (lower) the frequency bandwidth of the phase matching function is, the better the temporal (frequency) estimation of a parameter is. 

Let us now comment the sensitivity that could be achieved with such a  state using state-of-the-art photon sources. A Gaussian phase-matching function of SPDC process can be engineered in optical integrated circuit as in \cite{francesconi_engineering_2020,ferreri_spectrally_2020}, or by four-wave mixing in atomic media in the group delay regime where the projected pump field is transferred onto the bandwidth with a scaling factor $v_{g}$, which is the group speed. Typical value of $\sigma$ is $2\pi 10.9$ THz in AlGaAs optical integrated devices \cite{PhysRevA.102.012607}, or in the order of $0.1$ THz in \cite{francesconi_engineering_2020}. For periodically-poled  potassium  titanyl phosphate (PPKTP) crystal, it is possible to engineer the phase-matching function by modulating the duty cycle of the grating structure, which is centered at 1582 nm with a width of 5 nm (0.5 THz) in \cite{BenDixon:13}. This engineering was first pointed out in \cite{branczyk_engineered_2011}, where a Gaussian phase-matching function has been implemented with an indirect  modulation of the crystal nonlinearity. Such a technique, called Gaussian apodization is also applicable for KTP crystal \cite{van_der_meer_optimizing_2020}. For photons produced by SPDC with laser-cooled atomic system (see for instance \cite{zhao_shaping_2015,Zhao:14}), the typical frequency width is $\sigma \sim 10^{6}$ Hz. In Table.~\ref{tableaurecap}, we present the QCR bound for a given number of repetitions of experiment $N$ for all these sources with Gaussian phase-matching function.  From these references values, the aim is to engineer the phase matching function to beat these values, for these two ranges of parameters.

By adding a quadratic frequency phase, the wavefunction becomes: $\ket{\psi}=\iint d\omega_{s}d\omega_{i}  f_{+}(\omega_{+}) f^{\sigma}_{-}(\omega_{-}-\Delta)f^{iC}_{-}(\omega_{-}-\Delta)  \ket{\omega_{s}}\ket{\omega_{i}}$, where $f^{iC}_{-}(\omega_{-}-\Delta)=\text{exp}(\pm i(\omega-\Delta)^{2}/2C^{2})$. The corresponding chronocyclic Wigner distribution is:
\begin{equation}
W_-(\omega,t)= \text{exp}(-\frac{(\omega-\Delta)^{2}}{\sigma^{2}})\text{exp}(-t^{2}\sigma^{2})\text{exp}(\mp \frac{\omega t \sigma^{2}}{C^{2}}) .
\end{equation}
 Then the QFI has non zero anti-diagonal elements:
\begin{equation}\label{QFIsingle}
{\cal{F}}=\frac{1}{2}\begin{pmatrix}
\sigma^{2} & \mp \sigma^{2}/C^{2} \\ \mp \sigma^{2}/C^{2} & 1/\sigma^{2}+\sigma^{2}/C^{4}
\end{pmatrix}.
\end{equation}
 A chirp adds a time-frequency correlation ($\langle \omega_{-}t_{-} \rangle \neq 0$) and increases the temporal variance and can be created by engineering a temporal lens $C^{2}=40$ kHz/$\mu$s in \cite{mazelanik_temporal_2020}. In this last example, we point out that $\sigma^{2}/C^{4}=6.25 \ 10^{-10}\text{Hz}^{-2}$ while $1/\sigma^{2}\sim 10^{-12}\text{Hz}^{-2}$  then the frequency chirp used in laser-cooled atomic system is indeed beneficial for frequency estimation parameter. Then, the inverse of the QFI is:
 \begin{align}
 \tilde{{\cal{F}}}_{\tau\tau}=& \ 1/\sigma^{2}+\sigma^{2}/C^{4}\\
 \tilde{{\cal{F}}}_{\mu\mu}=&\  \sigma^{2}\\
 \tilde{{\cal{F}}}_{\mu\tau}=&\  \sigma^{2}/C^{2}.
 \end{align}
The corresponding QCR bound are indicated with a red color in Table.~\ref{tableaurecap}. The use of such a chirp does not improve temporal estimation but it is beneficial for simultaneous measurement of a time and frequency parameter. 
 Alternatively, by using a temporal chirp (noted $\tilde{C}$ in unit of time), the inverse of the QFI is:
   \begin{align}
 \tilde{{\cal{F}}}_{\tau\tau}=& \ 1/\sigma^{2}\\
 \tilde{{\cal{F}}}_{\mu\mu}=& \ \sigma^{2}+1/(\sigma^{2}\tilde{C}^{4})\\
 \tilde{{\cal{F}}}_{\mu\tau}=& \ \sigma^{2}\tilde{C}^{2}.
 \end{align}
Typical value of the chirp of a pump $\tilde{C}^{2}\sim 10^{5} fs^{2}$ in bulk platform is for instance \cite{brecht_characterizing_2013}, with $\sigma\sim 1.5$ THz, then $\sigma^{2}\gg  1/\sigma^{2}\tilde{C}^{4}$ which means that the typical value of temporal chirp is not useful for temporal estimation. The corresponding QCR bound are indicated with a blue color in Table.~\ref{tableaurecap}.  For integrated optics, such as  AlGaAs devices \cite{francesconi_engineering_2020}, the phase matching function of the photon pair is the spatial profile of the pump, the quadratic chirp could then be engineered with a spatial light modulator.

The expression of the FI can also be handled. In particular, the expression of the temporal FI for zero chirp $C=0$ has been given in \cite{lyons_attosecond-resolution_2018}. Note that the frequency FI has the same expression as the temporal one in that case, just by inverting $\sigma$ to $1/\sigma$ in the expression of interest.  When the frequency chirp is different than zero, $1/\sigma^{2}$ would be replaced by $1/\sigma^{2}+\sigma^{2}/C^{4}$, and it is then possible to reach a lower value of the Cramer-Rao bound. Finally, even if there are photodetectors losses, the choice of parameters that increase the QFI, also increase the FI.

\begin{table*}
    \begin{tabular}{ | l | l | l | p{1.4cm} |}
    \hline
     & $ \Delta\tau\sqrt{N}$ (s) & $ \Delta\mu\sqrt{N}$ (Hz) & $ \Delta(\mu\tau)\sqrt{N}$ \\ \hline
    AlGaAs device  \cite{PhysRevA.102.012607} & $10^{-14}$ & $10^{14}$ & / \\ \hline
    AlGaAs device \cite{Orieux:11} and PPKTP device \cite{BenDixon:13} & $10^{-12}$ & $10^{12}$ & / \\ \hline
    Laser Cooled atomic system \cite{zhao_shaping_2015,Zhao:14} & $10^{-6}$, \news{$10^{-5}$} & $10^{6}$, \news{$10^{6}$} & \news{25} \\   \hline
    Bulk system \cite{brecht_characterizing_2013} & $10^{-13}$, \bcomment{$10^{-13}$} & $10^{13}$, \bcomment{$10^{13}$} & \bcomment{$10^{9}$}\\
    \hline
    \end{tabular}
            \caption{\label{tableaurecap}Typical values of the QCR bound for Gaussian phase matching function, considering physical parameters of nowadays photon sources. In red, values reachable in our knowledge by using a frequency chirp for bulk system. In blue, value reachable in our knowledge with a temporal chirp for the laser cooled atomic platform. In the left column, we have also indicated examples of experimental set-ups. N stands for the number of repetitions of the experiment.}
\end{table*}

\section{Parameter estimation with cat-like states}\label{sectionthree}
In this section we analyze estimation precision achievable with time-frequency cat-like states~\cite{PhysRevA.102.023710}, which are biphoton states with a non-Gaussian chronocyclic Wigner function.
\subsection{Estimation of temporal or frequency shifts}\label{secduchat}
\begin{figure*}
\begin{center}
 \includegraphics[scale=0.18]{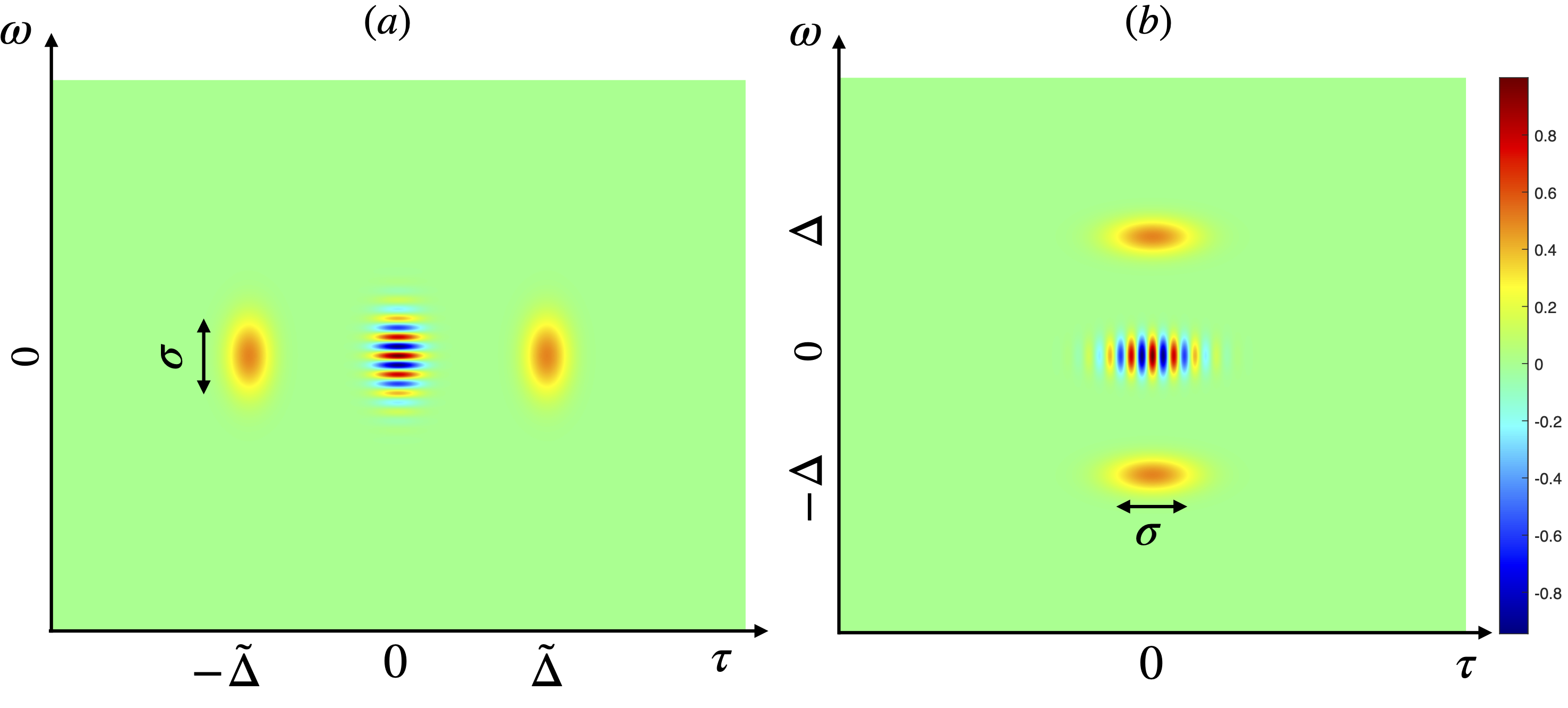}
 \caption{\label{wignercat} Chronocyclic Wigner distribution of the time cat-like state (left) and frequency cat-like state (right). On the left, the units of the horizontal axis (resp. vertical) is the $\mu$s (resp. MHz) for atomic laser-cooled system. On the right, the units of the horizontal (resp. vertical) axis is ps (resp. 1 THz) with the examples given bulk, but for the integrated sources it is $10^{-1}$ ps (resp. 100 GHz).     }
 \end{center}
\end{figure*}

The phase-matching function of a frequency cat-like state is given by the sum of two Gaussian functions centered at different frequencies. Accordingly, the chronocyclic Wigner function of a time-frequency cat-like state is of the same form of the phase-space Wigner distribution of a standard Schroedinger cat state.
In Ref.~\cite{PhysRevA.102.012607}, we explained why the time-frequency continuous variables of single photon are mathematically analogous to the quadrature position-momentum variables, owing to the underlying non-commutative algebra of creation and annihilation operators corresponding to frequency or arrival-time degrees of freedom. The wave function of the frequency cat-like state can be written as:
\begin{multline}\label{frequencycatstate}
\ket{\psi}=N_{+} \iint \text{d}\omega_{s} d\omega_{i} f_{+}(\omega_{+}) (f^{\sigma}_{-}(\omega_{-}-\Delta)+ f^{\sigma}_{-}(\omega_{-}+\Delta))\\
\cross \ket{\omega_{s}}\ket{\omega_{i}}\\=N_{+} (\ket{\psi_{\Delta}}+\ket{\psi_{-\Delta}}),
\end{multline}
where $N_{+}$ is a normalization factor given by $N_{+}^{2}(1+\text{exp}(-\Delta^{2})/\sigma^{2})=1$. The Fourier transform of the frequency cat-like state Eq.~(\ref{frequencycatstate}) is:
\begin{equation}\label{phasematchingcat}
\tilde{f}_{-}(t_{-})=2N_{+} \text{cos}(\Delta t_{-}) e^{-t_{-}^{2}\sigma^{2}}.
\end{equation}
The Chronocyclic Wigner distribution of the phase-matching function is:
\begin{multline}
W_- (\omega_{-},\tau)=N_{+}\big[e^{-\tau^{2}\sigma^{2}}e^{-(\omega_{-}+\Delta)^{2}/\sigma^{2}}+e^{-\tau^{2}\sigma^{2}}\\ \cross e^{-(\omega_{-}-\Delta)^{2}/\sigma^{2}}+2 e^{-\omega_{-}^{2}/\sigma^{2}} e^{-\tau^{2}\sigma^{2}}\text{cos}(2\Delta\tau)\big],
\end{multline}
and is represented in Fig.~\ref{wignercat} (b). The distribution is composed of two auto-terms, which correspond to the two peaks centered at $(\pm \Delta, 0)$ and an interference pattern along the time axis. The presence of the oscillations indicates that the state is indeed a coherent superposition of two Gaussian functions. We remind that the usual HOM experiment when the frequency shift is set to zero gives the coincidence probability $I(0,\tau)=1/2(1-W_{-}(0,\tau))$, which is represented in Fig.~\ref{FICAT} (a).

\begin{figure}[h]
\begin{center}
\includegraphics[scale=0.22]{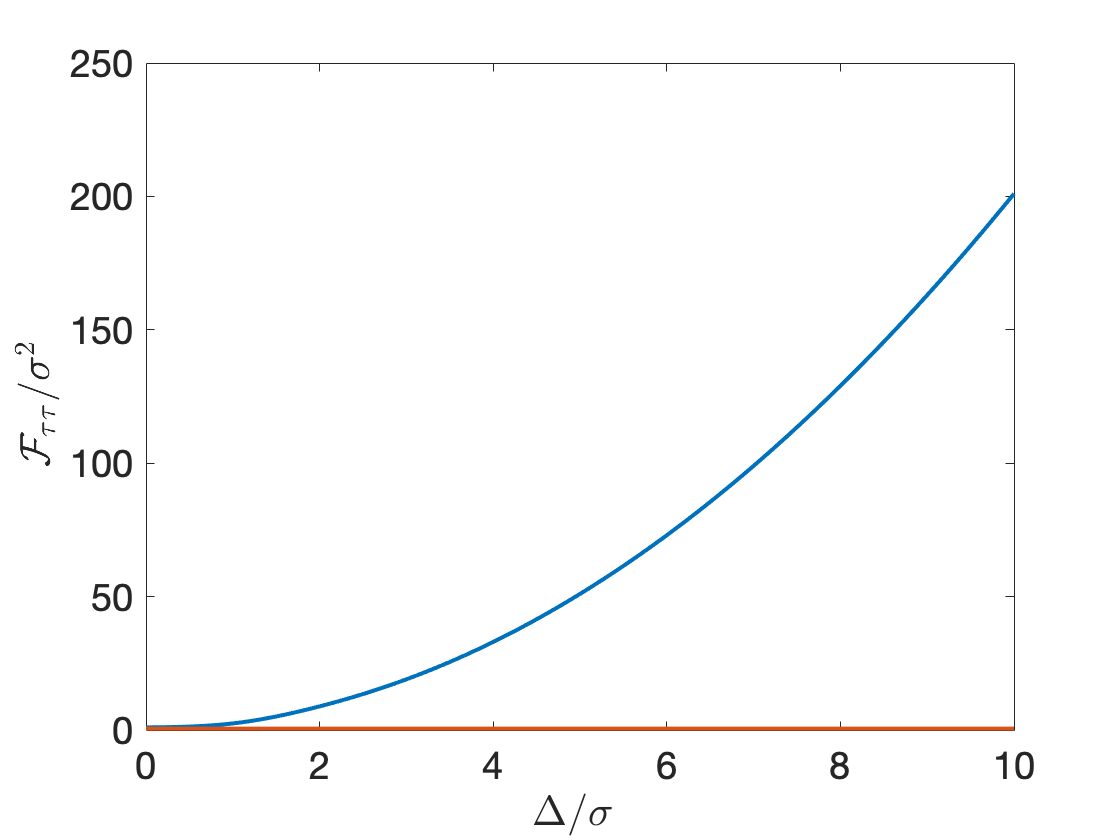}
 \caption{\label{QFICAT} Expression of the temporal QFI normalized by the width of each peak as function of the ratio $(\Delta/\sigma)$ for the frequency cat-like state (in blue) and for the time-frequency Gaussian state (in red).}
 \end{center}
\end{figure}

\begin{figure}[h]
\begin{center}
\includegraphics[scale=0.22]{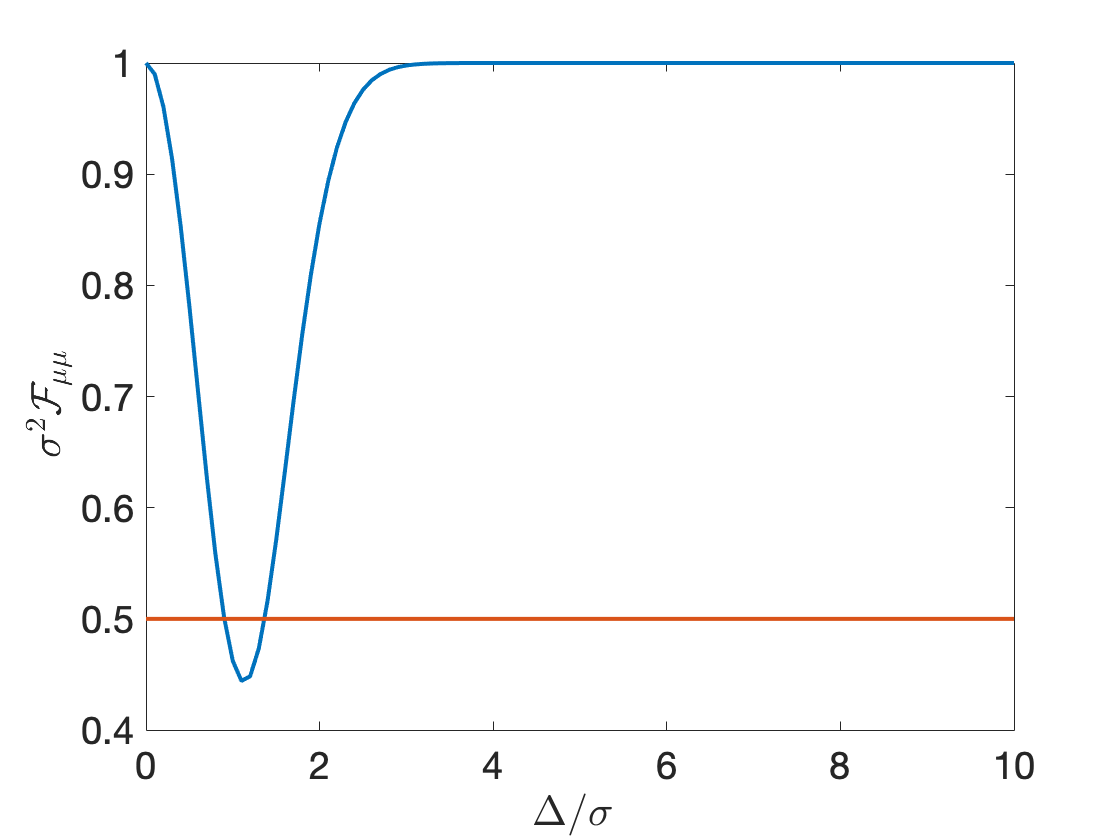}
 \caption{\label{QFICAT1}  Expression of the frequency QFI normalized by the width of each peak  as function of the ratio $(\Delta/\sigma)$ for the frequency cat-like state (in blue) and for the time-frequency Gaussian   (in red).}
 \end{center}
\end{figure}

An evaluation of the Quantum Fisher information for the frequency cat-like state gives,
\begin{align}
{\cal{F}}_{\tau\tau}=& \  \frac{\sigma^{2}(1+e^{-\Delta^{2}/\sigma^{2}})+2\Delta^{2}}{1+e^{-\Delta^{2}/\sigma^{2}}}=\sigma^{2}+\frac{2\Delta^{2}}{1+e^{-\Delta^{2}/\sigma^{2}}}, \label{QFItime}\\
{\cal{F}}_{\mu\mu}=& \  \frac{1}{\sigma^{2}}(1-2\frac{\Delta^{2}}{\sigma^{2}} \frac{e^{-\Delta^{2}/\sigma^{2}}}{1+e^{-\Delta^{2}/\sigma^{2}}})\label{temporalQFIcat},\\
{\cal{F}}_{\mu\tau}= &  \ 0.
\end{align}
These results are represented in Fig.~\ref{QFICAT} and \ref{QFICAT1}. The temporal QFI of the frequency cat-like state Eq.~(\ref{QFItime}) with different notation has been provided  in \cite{chen_hong-ou-mandel_2019}, while our formalism allows us to extend this result to multiparameter estimation of time and frequency displacements.  The QFI of the coherent superposition of frequency cat-like state is  indeed greater than the one of a coherent state $\sigma^{2}$. In general, the frequency peaks are  well separated, which means $\Delta=10\sigma$, and then the denominator of the normalization factor is equal to $1+e^{-\Delta^{2}/\sigma^{2}} \sim 1$. Hence, the variance $\Delta \tau$ is reduced by a factor 10 compared to the time-frequency Gaussian state. For instance, for the AlGaAs waveguide \cite{Orieux:11}, the typical separation between the peaks is 5 nm, while the width of each peak is $0.5$ nm.
The fact that the state is better suited for temporal estimation and not for frequency estimation is reminiscent of the Heisenberg's inequality. In addition, the  non-diagonal quantum Fisher matrix element is zero ${\cal{F}}_{\mu\tau}=0$, and so this state is not suited for simultaneous frequency and arrival-time measurement.


Note that a statistical mixture of two colors $\hat{\rho}=1/2(\ket{\psi_{\Delta}}\bra{\psi_{\Delta}}+\ket{\psi_{-\Delta}}\bra{\psi_{-\Delta}})$ gives a lower QFI than the state in Eq.~\eqref{frequencycatstate}, as we obtain
\begin{multline}\label{QFImixedfreqcat}
{\cal{F}}_{\tau\tau}(\tau)=8e^{-2\tau^{2}\sigma^{2}}(\Delta \text{sin}(\Delta \tau)+\tau\sigma^{2} \text{cos}(\Delta \tau))^{2}\\+8\tau^{2}\sigma^{4}e^{-2\tau^{2}\sigma^{2}}e^{-2\Delta^{2}/\sigma^{2}}.
\end{multline}
For mixed states the QFI is not given by the variance, the derivation is shown in the Appendix \ref{QFImixedstate}.
 The QFI of the mixed state depends on the parameter to estimate $\tau$ and is lower than the QFI of the pure state. The first term corresponds to the square of the derivative of the Wigner distribution, and it gives the structure of double peaks. As $\tau\rightarrow 0$, the QFI of this mixed state goes to 0, which is the main difference compared to the pure case.

It is possible to exchange the role of the frequency and arrival-time variables. We define a time cat-like state, the state corresponding to the $\pi/2$ rotation of the chronocyclic Wigner distribution of the frequency cat-like state (see right side Fig.~\ref{wignercat}), whose wavefunction is given by,
\begin{multline}\label{temporalcat}
\ket{\psi}=N_{+}\iint dt_{s}dt_{i} \tilde{f}_{+}(t_{+})(\tilde{f}_{-}^{1/\sigma}(t_{-}-\tilde{\Delta})+\tilde{f}_{-}^{1/\sigma}(t_{-}+\tilde{\Delta}))\\\cross  \ket{t_{s}}\ket{t_{i}},
\end{multline}
the normalization factor being  $N_{+}^{2}(1+\text{exp}(-\tilde{\Delta}^{2}\sigma^{2}))=1$, and where $\tilde{\Delta}=t_{1}-t_{2}$, as it is for the frequency cat-like state. The chronocyclic Wigner distribution of such a state is
\begin{multline}\label{timecatstate}
W_- (\omega_{-},\tau)=N_{+}(e^{-\frac{\omega_{-}^{2}}{\sigma^{2}}}e^{-(\tau-\tilde{\Delta})^{2}\sigma^{2}}+e^{-\frac{\omega_{-}^{2}}{\sigma^{2}}}e^{-(\tau+\tilde{\Delta})^{2}\sigma^{2}}\\+2 e^{-\omega_{-}^{2}/\sigma^{2}}e^{-\tau^{2}\sigma^{2}} \text{cos}(2\omega_{-}\tilde{\Delta})),
\end{multline}
and it is represented in Fig.~\ref{wignercat}(a). The QFI has the same form as for the frequency cat-like state and can be written as:
\begin{align}\label{frequencytwotimecat}
{\cal{F}}_{\mu\mu}=& \frac{1}{\sigma^{2}}+\frac{2\tilde{\Delta}^{2}}{1+e^{-\tilde{\Delta}^{2}\sigma^{2}}},\\
{\cal{F}}_{\tau\tau}=& \  \frac{1}{\sigma^{2}}(1-2\frac{\tilde{\Delta}^{2}}{\sigma^{2}} \frac{e^{-\tilde{\Delta}^{2}/\sigma^{2}}}{1+e^{-\tilde{\Delta}^{2}/\sigma^{2}}}).
\end{align}
These functions are represented respectively in Fig.~\ref{QFICAT}  and Fig.~\ref{QFICAT1} since the frequency and time variables have been exchanged.
Laser-cooled atomic systems represent the most suitable platform to generate broader temporal distributions of single-photon wave-packets to be applied in frequency parameter estimation protocols.  The frequency-estimation error is improved by a factor $10$ with respect to a Gaussian phase-matching function, by  assuming that the difference between the arrival times of the photon pairs can reach the value $\tilde{\Delta}\sim10\sigma$, which corresponds to two well-separated temporal wave packets.

\subsection{Fisher information for frequency and time cat-like states}
\begin{figure*}
\begin{center}
 \includegraphics[scale=0.20]{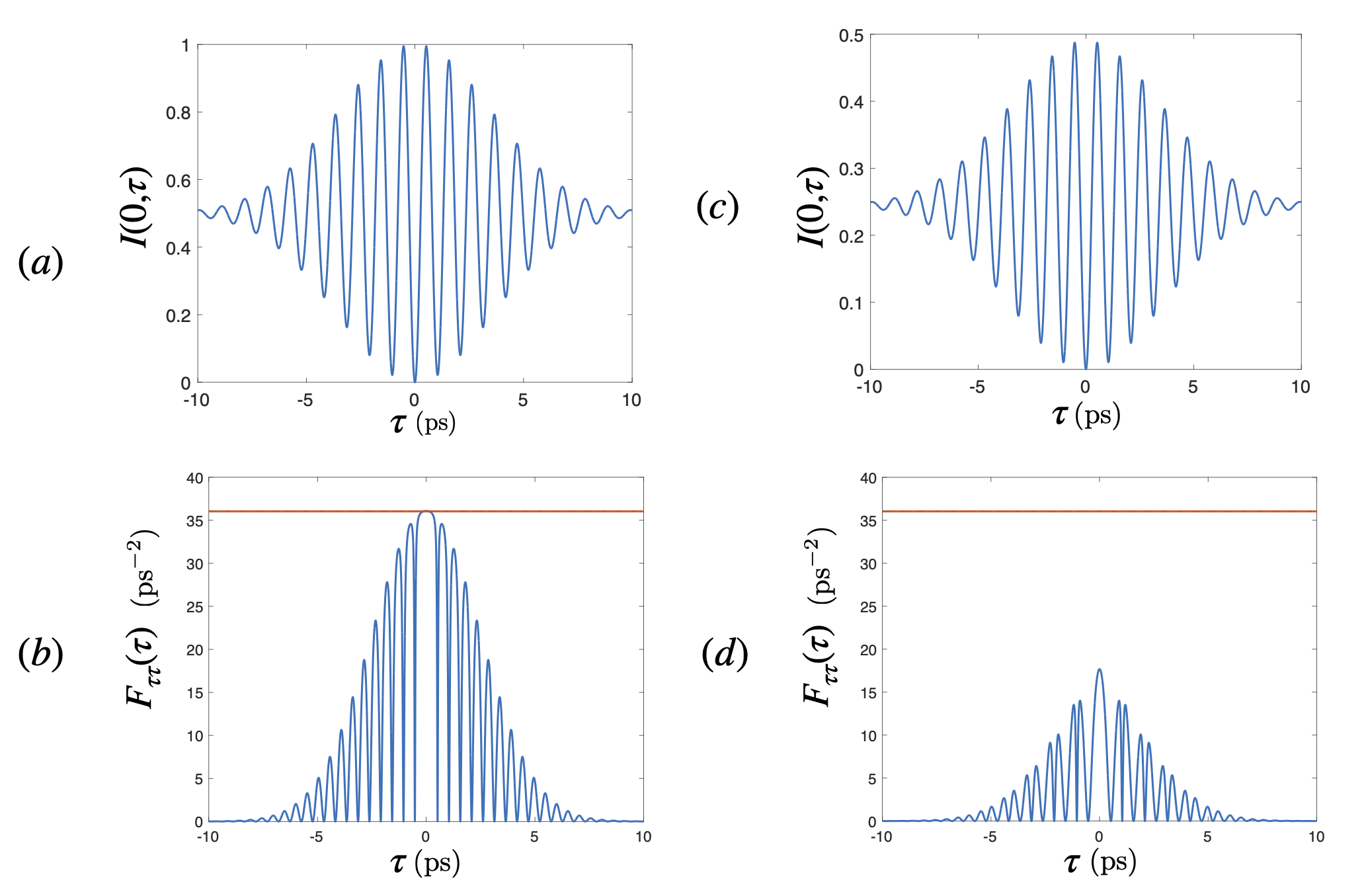}
 \caption{\label{FICAT}(a) (c) Coincidence probability $I(0,\tau)$ (see Eq.~(\ref{coincidenceprobability})) of the frequency cat-like state for $\Delta/\sigma=10$ and for different values of photodetector losses $\gamma=0,0.3$.  (b), (d) Expression of the FI as function of the time delay $\tau$ for the frequency cat-like state for different values of losses $\gamma=0,0.3$. The maximum of the FI is at $\tau=0$ but the FI does not reach the QFI in red (equals to $2\Delta^{2}+\sigma^{2}$) unless the photodetector losses are set to zero. The units depends on the used platform. We have chosen the units as  in the experimental platform \cite{chen_hong-ou-mandel_2019}.}
 \end{center}
\end{figure*}

The temporal Fisher information of the frequency cat-like state [see Eq.~\eqref{fisherinformationdeveloped}] can be cast as,
\begin{multline}\label{timeestimation}
F_{\tau\tau}(\tau)= \frac{1}{2}(1-\gamma)^{2}(2\Delta \text{sin}(2\Delta \tau)+\\ +  2\tau \sigma^{2}\text{cos}(2\Delta \tau))^{2}e^{-2\tau^{2}\sigma^{2}}\\
\times [\frac{1}{1-e^{-\tau^{2}\sigma^{2}}\text{cos}(2\Delta\tau)}+\frac{1}{\frac{1+3\gamma}{1-\gamma}+e^{-\tau^{2}\sigma^{2}}\text{cos}(2\Delta\tau)}]
\end{multline}
while for frequency estimation we obtain,
\begin{multline}\label{frequencyestimation}
F_{\mu\mu}(\mu)= \frac{1}{2}(1-\gamma)^{2}\frac{4}{N_{+}^{2}\sigma^{2}}((\mu+\Delta)e^{-(\mu+\Delta)^{2}/\sigma^{2}}\\
+(\mu-\Delta)e^{-(\mu-\Delta)^{2}/\sigma^{2}}+2\mu e^{-\mu^{2}/\sigma^{2}})^{2}\\
 \cross (\frac{1}{1-\frac{1}{N_{+}}(e^{-(\mu+\Delta)^{2}/\sigma^{2}}+e^{-(\mu-\Delta)^{2}/\sigma^{2}}+2e^{-\mu^{2}/\sigma^{2}}}\\+\frac{1}{\frac{1+3\gamma}{1-\gamma}+\frac{1}{N_{+}}(e^{-(\mu+\Delta)^{2}/\sigma^{2}}+e^{-(\mu-\Delta)^{2}/\sigma^{2}}+2e^{-\mu^{2}/\sigma^{2}})})).
\end{multline}
We represent  Eq.~(\ref{timeestimation}) in Fig.~\ref{FICAT} (b) and (d) for different values of the losses ($\gamma=0, 0.3$ and with $\Delta/\sigma=10$, with the corresponding coincidence probability $I(0,\tau)$ in Fig.~\ref{FICAT} (a) and (c). For instance, in Ref.~\cite{chen_hong-ou-mandel_2019}, $\sigma=250$ GHz and $\Delta=2.5$ THz, while in Ref.~\cite{francesconi_engineering_2020}, $\sigma\sim 50$ GHz and for a feasible spectral interval  $\Delta=500$ GHz. For zero losses, as $\tau\rightarrow 0$ and $\mu\rightarrow 0$, the Fisher information reaches the Quantum Fisher information (see the proof in the Appendix \ref{generalcasefisher}) because the phase matching is an even function, which means again that the measurement is optimal. The expression of Eq.~(\ref{frequencyestimation}) is given for the sake of completeness given that, even if it does not lead to a useful estimation strategy,  it shows the wide applicability of the chronocyclic Wigner function formalism. Notice that if the photons have another source of distinguishability as in \cite{chen_hong-ou-mandel_2019}, the FI is not maximum at $\tau=0$, even if there are no photodetector losses. In this case, the maximum of the FI must be found in a post-processing calculation, which allows finding the optimal point for measuring a time parameter.

Note that the temporal QFI Eq.~(\ref{timeestimation}) is equivalent to the frequency QFI of the time cat-like state Eq.~(\ref{temporalcat}). In other words, since the performance for temporal and frequency estimation is inverted when a $\pi/2$ rotation is processed, then the FI of the time cat-like state will also have the form Eq.~(\ref{timeestimation}), by replacing $\tau$ by $\mu$ and $\Delta$ by $\tilde{\Delta}$. Hence, the analysis presented in  \cite{chen_hong-ou-mandel_2019} holds for frequency estimation with a time cat-like state. The saturation of the Cramer-Rao bound can be done with the procedure explained in Appendix \ref{SaturationCramerRao}, in the case of simultaneous estimation of a frequency and time parameter.

\subsection{Effect of a temporal and frequency chirp}
Let us now discuss the addition of a chirp on the time or frequency cat-like state for simultaneous estimation. We can apply a frequency or a time chirp on a temporal or frequency cat-like state. They will both give the same off-diagonal elements of the QFI matrix. We then limit ourselves to present the case of the wave function of the time cat-like state. When a temporal chirp is added, the wave function is given by the convolutional product of the temporal spectrum Eq.~(\ref{temporalcat}), and the chirp function $g(t)=e^{-i\pi/4}e^{it^{2}C^{2}}$,
\begin{multline}\label{catstatechirp}
\ket{\psi}=N_{+}(\iint dt_{s}dt_{i} \tilde{f}(t_{+}) (\tilde{f}_{-}^{1/\sigma}(t_{-}-\tilde{\Delta})\circledast g(t_{-}-\tilde{\Delta})\\+\tilde{f}_{-}^{1/\sigma}(t_{-}+\tilde{\Delta})\circledast g(t_{-}+\tilde{\Delta})) \ket{t_s}\ket{t_i}),
\end{multline}
where $N_{+}$ is the normalization factor again given by  $N_{+}^{2}(1+\text{exp}(-\Delta^{2})/\sigma^{2})=1$. The associated chronocyclic Wigner distribution can be calculated (see for instance \cite{matz_wigner_2003}) as:
\begin{multline}
W_- (\omega_{-},\tau)=N_{+}(e^{-(\tau-\tilde{\Delta})^{2}\sigma^{2}}e^{-(\omega_{-}+(\tau-\tilde{\Delta})C^{2})^{2}/\sigma^{2}}\\+e^{-(\tau+\tilde{\Delta})^{2}\sigma^{2}}e^{-(\omega_{-}+(\tau+\tilde{\Delta})C^{2})^{2}/\sigma^{2}}\\+2  e^{-\tau^{2}\sigma^{2}}e^{-(\omega_{-}+\tau C^{2})^{2}/\sigma^{2}}\text{cos}(2\tilde{\Delta} \omega_{-})).
\end{multline}
The QFI for such a state is then:
\begin{align}
{\cal{F}}_{\mu\mu} =&\  \frac{1}{\sigma^{2}}+\frac{2\tilde{\Delta}^{2}}{1+e^{-\tilde{\Delta}^{2}\sigma^{2}}},\\
{\cal{F}}_{\tau\tau} =&\  \sigma^{2}+\frac{C^{4}}{\sigma^{2}}-\frac{\tilde{\Delta}^{2}\sigma^{4}}{2}\frac{e^{-\tilde{\Delta}^{2}\sigma^{2}}}{1+e^{-\tilde{\Delta}^{2}\sigma^{2}}}, \\
{\cal{F}}_{\mu\tau} =&\  \frac{(C/\sigma)^{2}}{1+e^{-\tilde{\Delta}^{2}\sigma^{2}}}.\label{multicat}
\end{align}
Then, the inverse of the QFI in the case where $\tilde{\Delta}\sigma\gg 1$:
\begin{align}
\tilde{{\cal{F}}}_{\mu\mu}=\frac{\sigma^{2}+C^{4}/\sigma^{2}}{1+2\tilde{\Delta}^{2}(\sigma^{2}+C^{4}/\sigma^{2})},\\
\tilde{{\cal{F}}}_{\tau\tau}=\frac{1/\sigma^{2}+2\tilde{\Delta}^{2}}{1+2\tilde{\Delta}^{2}(\sigma^{2}+C^{4}/\sigma^{2})},\\
\tilde{{\cal{F}}}_{\mu\tau}= \frac{(C/\sigma)^{2}}{1+2\tilde{\Delta}^{2}(\sigma^{2}+C^{4}/\sigma^{2})}.
\end{align}
The corresponding state with a Gaussian phase-matching function would give an off-diagonal QFI: $\tilde{{\cal{F}}}_{\mu\tau}=(C/\sigma)^{2}$. Hence, since $\tilde{\Delta}\sigma=10$, $\tilde{{\cal{F}}}_{\mu\tau}$ is reduced from a factor 100  with a cat-like state. For the other terms, their expressions show that in the same limit, there is no advantage for temporal and frequency estimation of using such a chirp.

The next step is to evaluate the Fisher information $F_{\mu\tau}(\mu,\tau)$. For the sake of brevity we do not show its expression here, the analysis is the same as for the temporal estimation using a frequency cat-like state. Since the phase-matching function is an even one, the FI converges to the QFI when $\gamma,\mu,\tau\rightarrow 0$. In such a limit, the generalized HOM interferometer allows one to fully benefit from the cat structure for simultaneous measurement of the time and frequency parameters.

\section{Parameter estimation with time-frequency grid states}\label{sectionfour}

In this section, we quantify the performance of estimating time and frequency parameters with a photon pair possessing a phase matching function with a periodic grid structure, using the time-frequency Gaussian state as a benchmark. We consider states that are generated when the non-linear medium producing a photon pair is placed inside an optical cavity. 
In this case, the Fabry-Perot structure of peaks \cite{PhysRevA.73.043806} will give the spectral distribution of the generated biphoton state.

\subsection{Quantum Fisher information for temporal grid states}

\begin{figure*}
\begin{center}
 \includegraphics[scale=0.26]{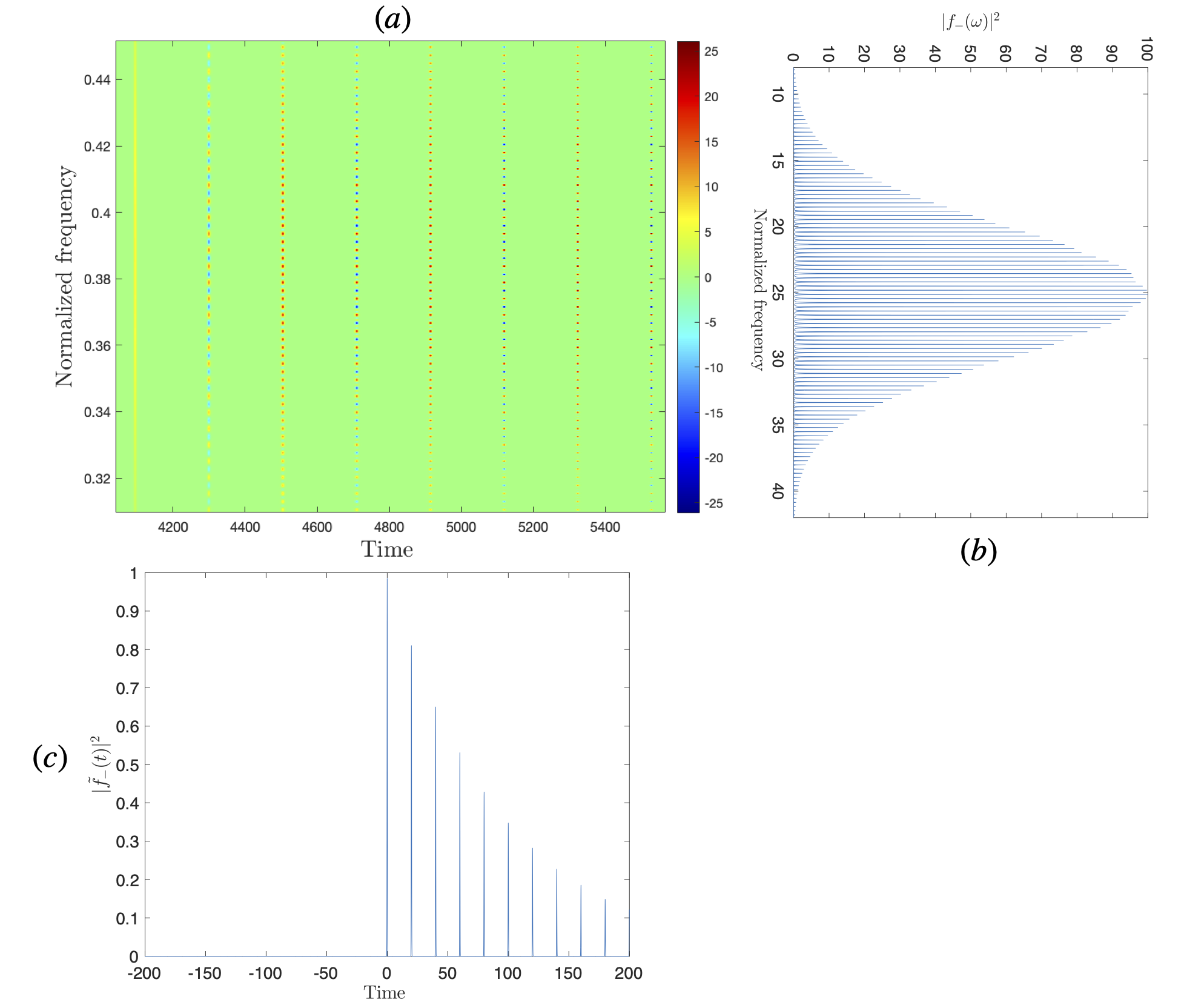}
 \caption{\label{wigner}  (a) Numerical Chronocyclic Wigner distribution $R=0.9$, $\sigma\overline{\tau}=20$. (b),(c) Marginals of the chronocyclic Wigner distribution, which are respectively the absolute value of the frequency and temporal phase matching function. The absence of reflection symmetry with respect to the origin to the temporal phase matching function is visible on the Wigner distribution. The peak on the left of (a) corresponds to the peak at zero time of the temporal phase matching function. The units of the chronocyclic Wigner distribution  are normalized with respect to the sampling frequency, while the ones of the marginals with respect to the periodicity of the comb. Concrete examples of units is given in Sec.~\ref{Fisherinformationgridssection}. }
 \end{center}
\end{figure*}

For anti-correlated photons, the frequency phase-matching function of what we will call a time-grid state can be written as:
\begin{equation}\label{phasefunction}
f_{-}(\omega_{-})=Ae^{-\frac{(\omega_{-}-\omega_{0})^{2}}{2\sigma^{2}}}T \sum_{n=0}^{\infty} R^{n} e^{2in(\omega_{-}-\omega_{0})\overline{\tau}},
\end{equation}
with the normalization factor $A$ given by $1=\abs{A}^{2} \sqrt{\pi\sigma^{2}}\sum_{n,m}R^{n+m}e^{-(n-m)^{2}\overline{\tau}^{2}\sigma^{2}}$. We denote with $\overline{\tau}=2\pi/\Delta\omega$ the round trip of the cavity,  being  $\Delta\omega$ the free spectral range. $T$ and $R$ are the transmission and the reflexion intensity coefficient of the optical cavity, respectively, while $\omega_{0}$ is the central frequency of the distribution. Beware that the sum is running over the positive integers, contrary to the Gaussian model (presented in the Appendix \ref{appendixgkp}). Such a function is represented in Fig.~\ref{wigner} (b) for $R=0.9$.  We define  the temporal phase-matching function as the Fourier transform of the phase matching function Eq.~(\ref{phasefunction}) is:
\begin{equation}
\tilde{f}_{-}(t_{-})=Ae^{i\omega_{0}t_{-}} T\sum_{n=0}^{\infty}R^{n}e^{-(t_{-}+2n\overline{\tau})^{2}\sigma^{2}/2},
\end{equation}
this function is represented in Fig.~\ref{wigner}(c). Note the absence of reflection symmetry with respect to the origin, which is a consequence of the sum running over the positive integer values. A frequency anti-correlated state can be produced by an AlGaAs optical integrated waveguide working at room temperature \cite{autebert_integrated_2016,PhysRevA.102.012607}. But many others experimental devices satisfy this condition, which justify the interest of the following application. Note that, we are dealing with time-frequency non-Gaussian state, such that even the envelop function is not necessary a Gaussian one, contrary to Sec.~\ref{QFIsinglecolorsection} where we only consider Gaussian phase matching function.

The chronocyclic Wigner distribution corresponding to Eq.~(\ref{phasefunction}) is given by
\begin{multline}\label{wignercombnormal}
W_{-}(\omega_{-},\tau)=\frac{{\cal{N}}}{\pi} \sum_{n,m} R^{n+m} e^{\frac{-(\omega_{-}-\omega_{0})^{2}}{\sigma^{2}}} e^{2i(n-m)(\omega_{-}-\omega_{0})\overline{\tau}} \\
\cross e^{-(\tau-(n+m)\overline{\tau})^{2}\sigma^{2}},
\end{multline}
which is represented in Fig.~\ref{wigner}(a) for $R=0.9$ and $\sigma\overline{\tau}=20$. The normalization factor in this case is given by ${\cal{N}}=\sum_{n,m}R^{n+m} e^{-\overline{\tau}^{2}\sigma^{2}(n-m)^{2}}$. The distribution is not symmetric along the two orthogonal directions, and we anticipate different performance for frequency and temporal parameter estimation. A coincidence measurement when the frequency shift is zero $I(\tau)$ corresponds to a measure of a cut of the Wigner distribution at zero frequency. All the peaks from this cut of the Wigner distribution are positive~\cite{PhysRevLett.91.163602,PhysRevA.102.012607}. 
Negative peaks of the Wigner distribution, which can lead to a coincidence probability larger than $1/2$, can be observed only with finite frequency shifts.
In the analytical expression of the chronocyclic Wigner distribution with the Airy function cavity model, the alternance of the positive and negative peaks is not analytically clear, as it is in the Dirac comb model. This last case was already studied \cite{testorf_fractional_1996} for a classical field and it also appears in \cite{gottesman_encoding_2001} for the calculation of the Wigner distribution of the quadrature position-momentum GKP state. More details are provided in Appendix (\ref{Diracmodel}).

For such a grid state Eq.~(\ref{wignercombnormal}), the quantum Fisher information matrix can be written as:
\begin{align}
{\cal{F}}_{\tau\tau}=&\ \frac{\sigma^{2}}{2}[1-2\overline{\tau}^{2}\sigma^{2}\text{Var}(n-m)],\label{fishercomb}\\
{\cal{F}}_{\mu\mu}=& \  \frac{1}{2\sigma^{2}}[1+2\overline{\tau}^{2}\sigma^{2}\text{Var}(n+m)]\label{frequencyyfishercomb},\\
{\cal{F}}_{\mu\tau}=& \ 0,
\end{align}
where the variance $\text{Var}(n\pm m)= \langle (n\pm m)^{2} \rangle-(\langle n\pm m \rangle)^{2}$  is taken with respect to:
\begin{equation}\label{averagevalue}
\langle (n\pm m)^{\alpha} \rangle= \frac{\sum_{n,m} (n\pm m)^{\alpha} R^{n+m}e^{-(m-n)^{2}\overline{\tau}^{2}\sigma^{2}} }{\sum_{n,m} R^{n+m} e^{-(m-n)^{2}\overline{\tau}^{2}\sigma^{2}}},
\end{equation}
where $\alpha=1,2$. The off-diagonal term of the QFI matrix ${\cal{F}}_{\mu\tau}$ can be different from zero if we add a quadratic phase in the phase-matching function as we shall see in Sec.~\ref{multiparameter}.  Note that the average $\langle n-m \rangle$ is equal to zero. Since there is a minus in the expression Eq.~(\ref{fishercomb}), there is no advantage to use such a state for temporal estimation, but for a certain parameter regime $R,\sigma,\overline{\tau}$, the variance can be negligible $\langle (m-n)^{2} \rangle-\langle n-m \rangle^{2}$. The fact that the precision of frequency estimation increases while the precision of the temporal estimation decreases is reminiscent of the Heisenberg inequality for time and frequency variables (see the case of the inequality based on the sum of variances \cite{PhysRevLett.113.260401}). The fact that such a grid state is useful for frequency estimation can be seen directly from the Chronocyclic Wigner distribution, which contains in the exponential imaginary linear terms in frequency but not in time variable.

\begin{figure*}
\begin{center}
 \includegraphics[scale=0.21]{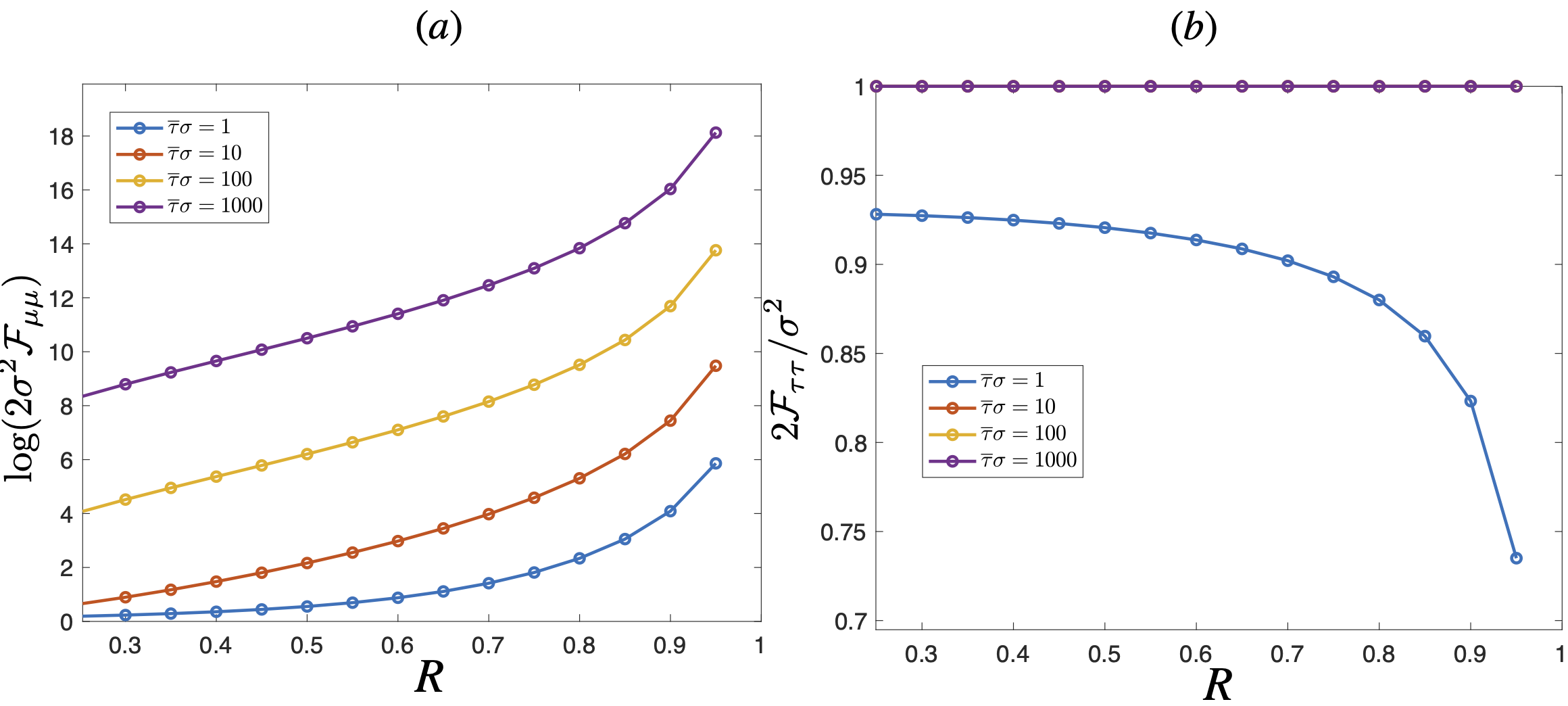}
 \caption{\label{frequencyesti} (a) Quantum Fisher information $2\sigma^{2}{\cal{F}}_{\mu\mu}=1+2\sigma^{2}\overline{\tau}^{2}\text{Var}(n+m)$ with respect to $R$ is the reflectivity of the cavity, for different value of the product $\sigma\overline{\tau}$. (logarithmic scale) (b) Quantum Fisher information $2{\cal{F}}_{\tau\tau}/\sigma^{2}=1-2\sigma^{2}\overline{\tau}^{2}\text{Var}(n-m)$ with respect to $R$ is the reflectivity of the cavity, for different values of the product $\sigma\overline{\tau}$. The red, yellow and purple curves coincide. }
 \end{center}
\end{figure*}

In Fig.~\ref{frequencyesti}(a), we represent $2\sigma^{2}{\cal{F}}_{\mu\mu}=[1-2\overline{\tau}^{2}\sigma^{2}\text{Var}(n-m)]$  for different values of the reflectivity and $\overline{\tau}\sigma$.  For simplicity, $\omega_{0}$ is set to zero. In general, as $R$ and $\overline{\tau}\sigma$ increase, the variance $\text{Var}(n+m)$ increases, and so it does the frequency QFI. The gain from the time-frequency grid state  through the factor $2\sigma^{2}\overline{\tau}^{2}\text{Var}(n+m)$ compared to the one gained by time cat-like state  $\frac{2\tilde{\Delta}^{2}}{1+e^{-\tilde{\Delta}^{2}/\sigma^{2}}}$ (see Eq.~\ref{frequencytwotimecat}) is remarkable and, as it can be seen in Fig.~\ref{frequencyesti}, it is relevant even for low reflectivity of the cavity mirrors.  In the experimental implementations~\cite{autebert_integrated_2016,PhysRevA.102.012607}, the free spectral range is $2\pi 19.2$ GHz, which gives the product $\sigma\overline{\tau}=3.51.10^{3}$, corresponding to approximately 500 frequency peaks.  In general, the product $\sigma\overline{\tau}$ approximately enumerates the number of peaks for each experimental platform. For instance, the device presented in \cite{imany_50-ghz-spaced_2018}, produced qudits with 40 frequency modes  and presents a good performance for the estimation of a frequency parameter according to Fig.~\ref{frequencyesti}(a). Also, the optical device in \cite{chang_648_2021}, producing 648 modes presents a great interest. 
On the contrary, the temporal QFI is lower than the time-frequency Gaussian state, and is represented in Fig.~\ref{frequencyesti}(b). When the temporal QFI is equal to one, there is no benefit of using such a time-frequency state compared to a Gaussian one. But then, as soon as $R\rightarrow 1$, the temporal QFI decreases, as a balance of the increasing of the frequency QFI.

Finally, we perform in the Appendix \ref{appendixgkp} the calculation of the QFI for a Gaussian micro-comb, which is the case in which the envelope is given by a Gaussian function. It allows one evaluating the performance of the time-frequency GKP state introduced in \cite{PhysRevA.102.012607}, which is actually quite similar to the state considered in this section, but with a different envelop function. In addition, this result could also be applied for quadrature position-momentum GKP states~\cite{gottesman_encoding_2001} and gives perspective for quantum metrology with a specific non-Gaussian quadrature state. This case is also relevant given that the phase-matching function is even, and so the FI can converge to the QFI.

\subsection{Quantum Fisher information for frequency grid states}\label{sectemporalcavity}
In order to outperform a Gaussian or a frequency cat-like state for temporal estimation, we have to invert the role of the frequency and the time parameters of the grid state Eq.~(\ref{phasefunction}). We call such a state a frequency grid state, since it is a $\pi/2$ rotation of the temporal grid state in the chronocyclic phase space. We propose to engineer the phase-matching function of the form:
\begin{equation}\label{temporalcavity}
\tilde{f}_{-}(t_{-})=e^{-t_{-}^{2}\sigma^{2}/2} \sum_{n=0}^{\infty}R^{n}e^{2int_{-}/\overline{\tau}},
\end{equation}
which can be seen as a frequency cat-like state (see Eq.~(\ref{phasematchingcat})) with a higher number of temporal peaks.  The associated chronocyclic Wigner distribution is indeed a $\pi/2$ rotation of Eq.~(\ref{wignercombnormal}), which corresponds to an exchange between the frequency and the time variable. In this case, we obtain opposite performances for frequency and time parameter estimation. Indeed, the QFI matrix is:
\begin{align}\label{QFIsecondgrid}
{\cal{F}}_{\mu\mu}=& \ \frac{1}{2\sigma^{2}}(1-2\overline{\tau}^{2}\sigma^{2}\text{Var}(n-m)),\\
{\cal{F}}_{\tau\tau}=& \ \frac{\sigma^{2}}{2}(1+ 2\overline{\tau}^{2}\sigma^{2}\text{Var}(n+m)),\\
{\cal{F}}_{\mu\tau}=& \ 0.
\end{align}
The QFI matrix for the state Eq.~(\ref{QFIsecondgrid}) is the same as in the previous section, we have just to exchange the role of ${\cal{F}}_{\mu\mu}$ by ${\cal{F}}_{\tau\tau}$ in Fig.~(\ref{frequencyesti}).
Such phase-matching function could be obtained by adding a SLM before a semiconductor AlGaAs device generating THz photon pairs \cite{francesconi_engineering_2020}, where the spatial pump profile $\tilde{f}$ is the phase-matching function $\tilde{f}(z)=\tilde{f}_{-}(t_{-}v_{g})$, where the scaling factor $v_{g}$ is the group speed velocity. We assume also that there is no cavity, and the full device and the detection is integrated, to avoid the natural Fabry-Perot effect when the semiconductor device is placed into the air [see Eq.~(\ref{phasefunction})].  In addition, for GHz photons produced by laser-cooled atoms system in the group delay regime, the phase matching function is also the spatial profile of the pump with the same rescaled group velocity factor and our results also apply for this measurement setup. 
\begin{figure*}
\begin{center}
 \includegraphics[scale=0.22]{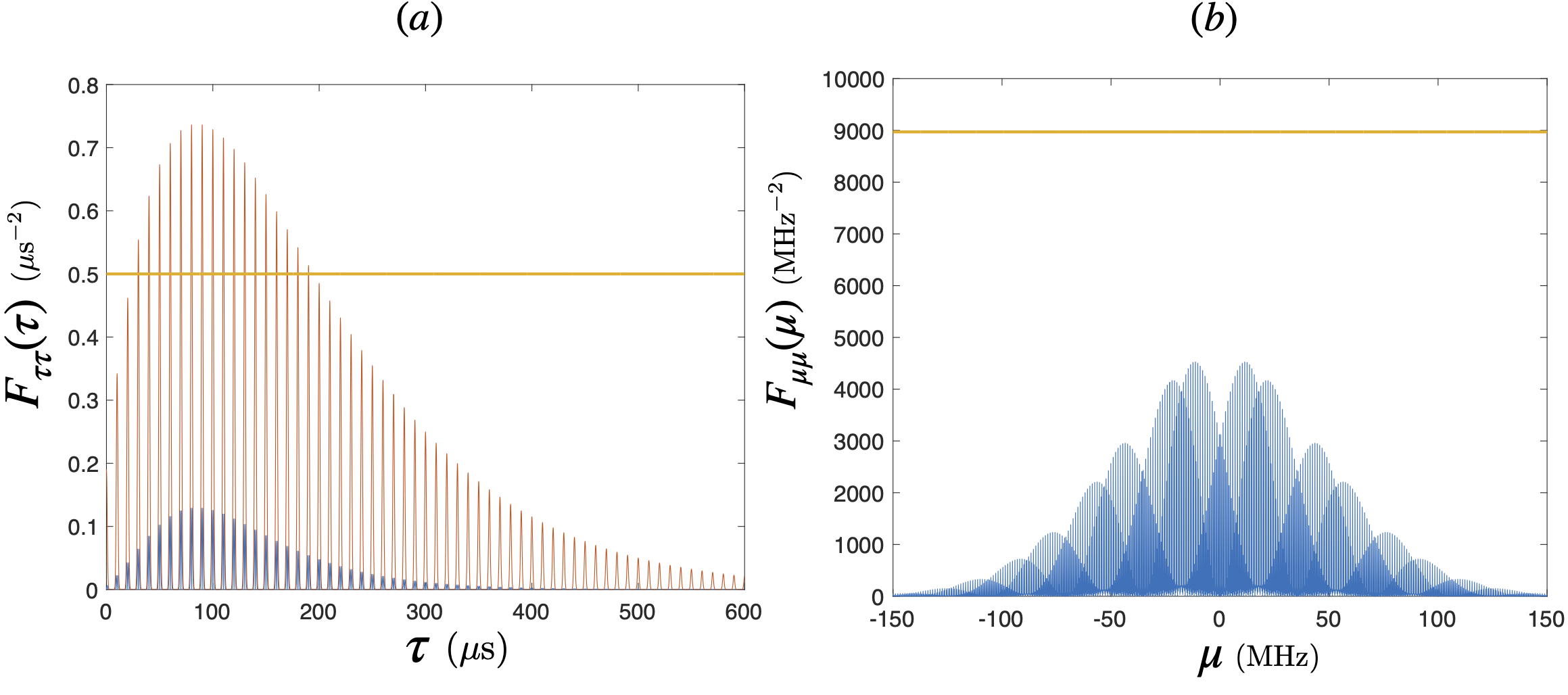}
 \caption{\label{classicalfishergrid} Metrological performance of Eq.~(\ref{fishercomb}),  with $R=0.9$ and $\overline{\tau}\sigma=10$. (a) In blue, analytical temporal Fisher information $F_{\tau\tau}(\tau)$ (in $\mu\text{s}^{-2}$ for atomic cooled laser system) for $\gamma=0$ (no losses) with respect to the temporal delay (in $\mu$s). In red, Cut of the Wigner distribution at zero frequency (proportional to the coincidence probability of the HOM experiment).  In yellow, QFI of the grid state Eq.~(\ref{fishercomb}).   (b) In blue, analytical frequency Fisher information $F_{\mu\mu}(\mu)$ ($\text{MHz}^{-2}$) for $\gamma=0$ (no losses) with respect to the frequency delay (MHz). In yellow, QFI of the grid state (see Eq.~(\ref{frequencyyfishercomb})) which is equal to 9000 $\text{MHz}^{-2}$ . Even if the FI does not converge towards the QFI, we gain an huge factor compare to the cat-like state. By figure of comparison, the QFI of the time cat-like state is 40 for the parameters $\tilde{\Delta}\sigma \sim 10$ for the same total bandwidth with the time-frequency grid state. }
 \end{center}
\end{figure*}

\subsection{Fisher information matrix for time-frequency grid states}\label{Fisherinformationgridssection}
The Fisher information for the estimation of a temporal parameter with the grid state Eq.~(\ref{phasefunction}) $F_{\tau\tau}(\tau)$ is represented in Fig.~\ref{classicalfishergrid}(a), where we have chosen $\overline{\tau}\sigma=10$ and $R=0.9$. In order to compare with the performance of the frequency (or time) cat-like state described in  \cite{chen_hong-ou-mandel_2019}, we can choose $\sigma=1$ THz so that to have the same total bandwidth corresponding to the spectral separation of the two frequencies of the frequency cat-like state.  The round-trip of the cavity will be then $\overline{\tau}\sim 10$ ps. 
The Fisher information presents a periodic structure with a maximum at $\overline{\tau}=100$ps, maximum which depends on the number of modes and the reflectivity $R$. The presence of double peaks in the FI is due to the presence of the derivative of the cut of Wigner distribution, which is also the case for the cat-like state. Note that, the FI does not saturate the QFI [the yellow line in Fig.~\ref{classicalfishergrid}(a)] because the phase-matching function is not even. As $\gamma$ increases, the entire curve shrinks. The choice of the value of the parameter $\overline{\tau}\sigma=10$ and $R=0.9$ is taken to show that there is no need to have a very large number of frequency peaks (or in other words entangled qudits) to achieve good performance for frequency estimation.

The Fisher information for the estimation of a frequency parameter with the grid state Eq.~(\ref{phasefunction})  is given in Fig.~\ref{classicalfishergrid}(b), for $\gamma=0$. Again, the FI does not reach the QFI even for zero photodetector losses. The optimal points for frequency estimation are slightly on the left or the right of the origin and can be found numerically after the measurement of the coincidence and the single count. Even if the FI does not converge to the QFI, we obtain better performance compared to what it is possible to achieve with reasonable parameters with a time cat-like state  $\sigma\tilde{\Delta}=10$ (see Sec.~\ref{secduchat})  and with $\tilde{\Delta}$ equals to the temporal envelope width of the grid state. Indeed, since we have fixed the number of photons, we compare the metrological performance of different biphoton spectral distributions according to their total bandwidth. Then, there is a gain for the chosen parameters in the FI of approximately $10^{2}$ (see Fig.~\ref{frequencyesti}), which leads to an improvement of the frequency resolution by a factor 10. Notice that by using the grid state Eq.~(\ref{temporalcavity}), the graph Fig.~\ref{classicalfishergrid} (a) and (b) are inverted, a relevant situation for both bulk and integrated optical devices. By consequence, the estimation of the temporal parameter would be sub-attosecond.  However, we considered conservative physical parameters, a higher reflectivity of the cavity and higher product $\overline{\tau}\sigma$ would lead to better performance for both frequency (with state Eq.~(\ref{phasefunction}) ) and temporal estimation (with state Eq.~(\ref{temporalcavity})). 
We point out that for $\gamma=0$, we have $\delta\alpha=1/\sqrt{F_{\mu,\tau}}$ (see Eq.~(\ref{timedelay})), which means that we saturate the Cramer-Rao bound. Besides, if there is a source of indistinguishability of the photon pair, the   maximum of the FI is no longer at the points indicated in Fig.~\ref{classicalfishergrid}(b).
Again, for time parameter estimation,  the state Eq.~(\ref{temporalcavity}) is the most suitable.

\begin{table}[h]
    \begin{tabular}{ | l | l | l | p{1.5cm} |}
    \hline
     & $ \Delta\tau\sqrt{N}$ (s) & $ \Delta\mu\sqrt{N}$ (Hz) & $ \Delta\mu\tau\sqrt{N}$ \\ \hline
    Gaussian state& $10^{-14}$ & $10^{6}$ & 25 \\ \hline
    Cat-like state & $10^{-15}$ & $10^{5}$ & 2.5 \\ \hline
    Grid state & $[10^{-14},10^{-17}]$ & $[10^{3},10^{6}]$ & $[10^{-2},10^{1}]$ \\
       \hline
    \end{tabular}
            \caption{\label{tableaucompaallstate} For a fixed total bandwidth, comparison of the reachable temporal and frequency estimation for the Gaussian, time or frequency cat-like state and  the time and frequency grid state. The bracket in the grid state line refers to all possible values obtained by varying $R$ and $\tau$, according to Fig.~\ref{frequencyesti}. N is the number of repetitions of the experiment.}
\end{table}

\subsection{Frequency chirp and multiparameter estimation with temporal grid states}\label{multiparameter}

In this section, we analyze the QFI for grid states when a chirp is applied. In particular, we consider a quadratic temporal chirp on a frequency grid state and, respectively, a frequency chirp on a temporal grid state.  Let us consider a phase-matching function which has a grid structure (see Eq.~(\ref{phasefunction})) but with an additional frequency chirp:
{\small{\begin{equation}\label{phasefunctionchirp}
f_{-}(\omega_{-})=Ae^{-\frac{(\omega_{-}-\omega_{0})^{2}}{2\sigma^{2}}}e^{-i\frac{(\omega_{-}-\omega_{0})^{2}}{2C^{2}}}T \sum_{n=0}^{\infty} R^{n} e^{2in(\omega_{-}-\omega_{0})\overline{\tau}},
\end{equation}}}
where $C$ is the frequency chirp parameter as in Sec.~\ref{QFIsinglecolorsection}.  The chronocyclic Wigner distribution is this case can be cast under the form
\begin{multline}\label{frequencyphasewigner}
W_-(\omega,\tau)=\frac{{\cal{N}}}{\pi} \sum_{n,m=0}^{\infty} R^{n+m} e^{-(\omega-\omega_{0})^{2}/\sigma^{2}} e^{2i(n-m)(\omega-\omega_{0})\overline{\tau}} \\
\cross e^{-(\tau-(n+m)\overline{\tau}-\frac{(\omega-\omega_{0})}{C^{2}})^{2}\sigma^{2}}.
\end{multline}
where the normalization constant is again given by ${\cal{N}}=\sum_{n,m}R^{n+m} e^{-\overline{\tau}^{2}\sigma^{2}(n-m)^{2}}$. The frequency $\omega_{0}$ will be set to zero for simplicity. The QFI matrix is given by (see Appendix \ref{calculationvariance}),
\begin{align}
\label{multiparamatrix}
{\cal{F}}_{\mu\mu}=& \frac{1}{2\sigma^{2}}(1+2\overline{\tau}^{2}\sigma^{2}\text{Var}(n+m)) \nonumber \\
+&\frac{\sigma^{2}}{2C^{4}}(1-2\overline{\tau}^{2}\sigma^{2}\text{Var}(n-m)),\\
{\cal{F}}_{\tau\tau}=& \ \frac{\sigma^{2}}{2}(1- 2\overline{\tau}^{2}\sigma^{2}\text{Var}(n-m)),\\
{\cal{F}}_{\mu\tau}=& \frac{\sigma^{2}}{2C^{2}}(1-2\overline{\tau}^{2}\sigma^{2}\text{Var}(n-m)).
\end{align}

This expression is composed of the same terms as the one of the time-frequency Gaussian state Eq.~(\ref{QFIsingle}) but with different multiplicative factors coming from the grid state. The inverse of the QFI matrix is given by:
\begin{align}
\tilde{{\cal{F}}}_{\mu\mu}=& \frac{2\sigma^{2}}{1+2\overline{\tau}^{2}\sigma^{2}\text{Var}(n+m)},\\
\tilde{{\cal{F}}}_{\tau\tau}=& \frac{2/\sigma^{2}}{1-2\overline{\tau}^{2}\sigma^{2}\text{Var}(n-m)}+\frac{2\sigma^{2}/C^{4}}{1+2\overline{\tau}^{2}\sigma^{2}\text{Var}(n+m)},\\
\tilde{{\cal{F}}}_{\mu\tau}=& -\frac{2\sigma^{2}/C^{2}}{1+2\overline{\tau}^{2}\sigma^{2}\text{Var}(n+m)}.
\end{align}
With the parameters  $R=0.9$, $\overline{\tau}\sigma\gg 10$ the variance $\text{Var}(n-m)\sim 10^{-44}$ is negligible. The inverse of the temporal QFI is in practice the same as if there was no chirp $\tilde{{\cal{F}}}_{\tau\tau}=2/\sigma^{2}$. The grid structure of the state allows one to drastically reduce $\tilde{{\cal{F}}}_{\mu\tau}$ compared to the Gaussian case through the term $1+2\overline{\tau}^{2}\sigma^{2}\text{Var}(n+m)\sim 500$ according to Fig.~\ref{frequencyesti}(a) where $\overline{\tau}\sigma=10$, for instance.  In Table.~\ref{tableaucompaallstate}, we finally recap the temporal and frequency precision reachable by the state presented in this paper.

\section{Conclusions}
\label{conclusions}

We have analyzed the performance of frequency and temporal parameter estimation protocols that make use of photon pairs as probe states, and which are based on generalized HOM interferometry. 
In particular, we have compared various classes of phase-matching functions that can be engineered with nowadays technology, considering continuous variables defined as the frequency and arrival time of biphoton states. Our results show that the estimation precision can be substantially improved by engineering time-frequency grid states, whose chronocyclic Wigner function is composed of a superposition of equally-spaced peaks.  We have also shown that the optimal limit of precision, the quantum Cramer-Rao bound, can be achieved using a HOM measurement on states with even phase-matching functions. A quantitative estimation of the precision achievable with state-of-the-art devices is presented, comparing the expected performances of specific sources of biphoton states. Our results pave the way to novel applications of quantum photonic devices in sensing applications, in particular in cases where intense light fields can not be used, such as in biological sensing~\cite{TAYLOR20161}.

Our results have been obtained working with the chronocyclic Wigner function formalism, but they could be generalized to other continuous variable systems. Indeed, the considered time-frequency grid state are the analogue of the Gottesman, Kitaev and Preskill states \cite{gottesman_encoding_2001}, which can be applied for displacement estimation~\cite{PhysRevA.95.012305,LEGERO2006253} and for continuous-variable quantum error correction~\cite{barros_free-space_2017,PhysRevA.93.012315,farias_quantum_2015}. 

As a future perspective, the non-diagonal elements of the QFI matrix, relevant for multiparameter estimation, could be improved generalizing our results to include continuous rotations in the chronocyclic phase space. Furthermore, other degrees of freedom such as polarization~\cite{dambrosio_photonic_2013} could be simultaneously exploited to enhance the measurement precision. Finally, multiphoton interference could be used in order to combine in a single protocol the  improvement coming from the engineering of the phase-matching function, with the metrological advantage coming from multi-photon entanglement.

\section*{ACKNOWLEDGMENT}
 N.Fabre acknowledges support from the project “Quantum Optical Technologies” carried out within the International Research Agendas programme of the Foundation for Polish Science co-financed by the European Union under the European Regional Development Fund. We acknowledge useful discussion with Perola Milman, Arne Keller, Sara Ducci, Florent Baboux, Arnault Raymond, Louis Garbe and Roberto Di Candia.

\appendix

\section{Quantum Fisher information}\label{fulldemonstrationQFI}

In this section, we present the expression of the QFI for a frequency anticorrelated biphoton state.
For a pure state the QFI  is given by ${\cal{F}}^{\text{tot}}_{\mu\tau}=4\left(\bra{\frac{\partial \psi(\tau)}{\partial \tau}}\ket{\frac{\partial \psi(\mu)}{\partial \mu}}-\abs{\bra{\psi(\mu)}\ket{\frac{\partial \psi(\mu)}{\partial \mu}}\bra{\psi(\tau)}\ket{\frac{\partial \psi(\tau)}{\partial \tau}}}\right)$, where the wavefunction of the biphoton state is written as: $\ket{\psi(\mu,\tau)}={\cal{\hat{D}}}(\tau)\hat{D}(\mu) \ket{\psi}=\iint d\omega_{s} d\omega_{i} e^{i \omega_{i}\tau} f_{+}(\omega_{+}+\mu) f_{-}(\omega_{-}+\mu) \ket{\omega_{s},\omega_{i}}$.
The QFI for temporal parameter estimation is,
\begin{multline}
{\cal{F}}^{\text{tot}}_{\tau\tau}=4(\iint d\omega_{s}d\omega_{i} \omega_{i}^{2} \abs{f_{+}(\omega_{+})}^{2}\abs{f_{-}(\omega_{-})}^{2}-\\
\abs{\iint d\omega_{s}d\omega_{i} \omega_{i} \abs{f_{+}(\omega_{+})}^{2}\abs{f_{-}(\omega_{-}})^{2}}^{2}),
\end{multline}
the temporal QFI can be written as the variance with respect to the frequency $\omega_{i}$ variable:
\begin{equation}
{\cal{F}}^{\text{tot}}_{\tau\tau}=4(\langle \omega^{2}_{i} \rangle-\langle \omega_{i} \rangle^{2})=4\text{Var}(\omega_{i}).
\end{equation}
 The average is evaluated with respect to the chronocyclic Wigner distribution of the photon pair:  $\langle \omega^{2}_{i} \rangle= \iint \iint d\omega_{s}d\omega_{i} dt_{s} dt_{i}W_{\hat{\rho}}(\omega_{s},\omega_{i},t_{s},t_{i}) \omega_{i}^{2}$. By performing the change of variable $\omega_i=(\omega_{+}-\omega_{-})$ and assuming a factorization as in Eq.~(\ref{factorization2}), the temporal QFI becomes:
\begin{equation}
{\cal{F}}^{\text{tot}}_{\tau\tau}=4(\langle (\omega_{+}-\omega_{-})^{2} \rangle- \langle  \omega_{+}-\omega_{-} \rangle^{2})= 4(\text{Var}(\omega_{+})+\text{Var}(\omega_{-})).
\end{equation}
For the other elements of the QFI matrix, we have an analogous expression: 
\begin{align}
{\cal{F}}^{\text{tot}}_{\mu\mu}=4 (\langle t_{s}^{2} \rangle- \langle t_{s} \rangle^{2})=4(\text{Var}(t_{+})+\text{Var}(t_{-}))  \\
{\cal{F}}^{\text{tot}}_{\tau\mu}=4( \langle \omega_{s} t_{i} \rangle - \langle \omega_{s} \rangle \langle t_{i} \rangle)=-4(\langle \omega_{-}t_{-}\rangle-\langle \omega_{-}\rangle \langle t_{-}\rangle),
\end{align}
We now mention two particular cases where the QFI matrix element depends only to the phase-matching part. The first is the case of frequency anti-correlated photon $W_{+}(\omega_{+},t_{+})=e^{-\omega^{2}_{+}/\sigma^{2}}e^{-t_{+}^{2}\sigma^{2}}$ where $\text{Var}(\omega_{-})\gg \sigma$, the QFI reduces to:
\begin{align}
 {\cal{F}}^{\text{tot}}_{\mu\mu}=4(\frac{1}{2\sigma^{2}}+ \text{Var}(t_{-}))\label{frequencyQFIanti} \\
 {\cal{F}}^{\text{tot}}_{\tau\tau}=4(\sigma^{2}+\text{Var}(\omega_{-})\sim \text{Var}(\omega_{-})).
\end{align}
In this case, the resolution of a temporal parameter depends only on the phase-matching function $f_{-}$ of the biphoton wavefunction. When the photodetectors losses are set to zero and when the phase-matching function is even, for a delay converging towards zero, the FI obtained from the HOM interferometer converges to the QFI. Nevertheless, the factor $1/2\sigma^{2}$ in the frequency QFI Eq.~(\ref{frequencyQFIanti}) is an improvement of the frequency estimation which is not reachable with the HOM interferometer. The second case is for frequency correlated photon pair, where the role of the frequency and the time variable can be inverted. 
Notice that  the moments are written as a statistical average, but it is also possible to write them in terms of quantum average with time and frequency operators by using the De-Wooters theory. It is possible to define such operators by adding a quantum clock as explained in \cite{maccone_quantum_2020,fabre:tel-03191301}.

\section{Optimal measurement strategy}\label{generalcasefisher}
In this section, we show that the FI for the HOM interferometer converges to the QFI when the photon detectors losses are zero and when the displacement parameters converge to zero. For simplicity, we consider here only temporal parameter estimation. First, we write the expression of the FI for lossless detectors,
\begin{equation}
F_{\tau\tau}(\tau)=\frac{1}{2} [ \frac{ (\partial_{\tau}W_{-}(\tau))^{2}}{1-W_{-}(\tau)}+\frac{ (\partial_{\tau}W_{-}(\tau))^{2}}{1+W_{-}(\tau)}].
\end{equation}
We perform a Taylor expansion of the cut of the Wigner distribution $W_{-}(0,\tau)\equiv W_{-}(\tau)$ at zero frequency and zero delay,
\begin{eqnarray}
W_{-}(\tau)=W_{-}(\tau=0)+\partial_{\tau}W_{-}(\tau=0)\tau+ \nonumber \\ + \partial^{2}_{\tau}W_{-}(\tau=0)\frac{\tau^{2}}{2}+O(\tau^{2})
\end{eqnarray}
For an even phase-matching function $f_{-}(\omega_{-})=f_{-}(-\omega_{-})$, we obtain that $\partial_{\tau}W_{-}(\tau=0)=2i\langle \omega_{-} \rangle$, where the average value is again taken with respect to $W_-$ (or after integration on the time variable, with respect to $\abs{f_{-}}^{2}$). Also, we have $W_{-}(\tau=0)=1$, owing to the normalization condition of the biphoton wavefunction and because the phase matching is an even function.  Then we obtain the following expression for different terms which intervene in the expression of the FI:
\begin{align}
1-W_{-}(\tau)= -\sum_{k=1}^{\infty} \frac{(2i\tau)^{k}}{k!} \langle \omega^{k} \rangle, \\
1+W_{-}(\tau)=2+ \sum_{k=1}^{\infty} \frac{(2i\tau)^{k}}{k!} \langle \omega^{k} \rangle,\\
\partial_{\tau} W_{-}(\tau)= \sum_{k=1}^{\infty} \frac{(2i)^{k}\tau^{k-1}}{(k-1)!} \langle \omega^{k} \rangle
\end{align}
For an even state centered at the origin, we have that $\langle \omega_{-} \rangle =0$, and so at the first order in $\tau$
\begin{align}
\frac{(\partial_{\tau} W_{-}(\tau))^{2}}{P_{2}(\tau)}=8\langle \omega_{-}^{2} \rangle,\\
\frac{(\partial_{\tau} W_{-}(\tau))^{2}}{P_{1}(\tau)}\rightarrow 0.
\end{align}
Then, the FI converges to $4\langle \omega_{-}^{2} \rangle$ when $\tau\rightarrow 0$ which is indeed the QFI $\mathcal F$ which depends only on the phase-matching function. The previous mathematical proof can also be applied to show that the frequency FI for $\gamma,\mu\rightarrow 0$, converges to the frequency QFI, by using the fact that the Fourier transform of a complex even function is also complex and even. The frequency cat-like state, time cat-like state with or without chirp and also for the Gaussian comb (see Appendix \ref{appendixgkp}) has indeed an even  phase-matching function. It is not the case for the Airy function model (see Eq.~(\ref{phasefunction})), since we have:
\begin{equation}
f_{-}(\omega_{-})\pm f_{-}(-\omega_{-})=\left\{
    \begin{array}{ll}
        e^{-\omega_{-}^{2}/\sigma^{2}} \frac{2-2R\text{cos}(\omega_{-}\tau)}{1-2R\text{cos}(\omega_{-}\tau)+R^{2}} \\
        e^{-\omega_{-}^{2}/\sigma^{2}} \frac{2i\text{sin}(\omega_{-}\tau)}{1-2R\text{cos}(\omega_{-}\tau)+R^{2}}
    \end{array}
\right.
\end{equation}
which is not an even or an odd function. In such a case, a different interferometric scheme must be devised to obtain an optimal measurement that saturates the QFI.

\section{QFI of mixture of two-colors state}\label{QFImixedstate}

We start by the coherent mixture of a biphoton state in two frequencies:
\begin{equation}
\hat{\rho}=\frac{1}{2}(\ket{\psi_{\Delta}}\bra{\psi_{\Delta}}+\ket{\psi_{-\Delta}}\bra{\psi_{-\Delta}}).
\end{equation} 
After time and frequency displacement, the state is: $\hat{\rho}(\mu,\tau)=\hat{D}(\tau){\cal{\hat{D}}}(\mu) \hat{\rho} \hat{\hat{D}}^{\dagger}(\tau){\cal{\hat{D}}}^{\dagger}(\mu)$.  The QFI matrix can be written~\cite{liu_quantum_2020} using the spectral decomposition of the density matrix $\hat{\rho}=\sum_{i=1}^{d} \lambda_{i}\ket{\lambda_{i}}\bra{\lambda_{i}}$ as,
\begin{equation}
{\cal{F}}_{\mu\tau}=\sum_{i,j=0}^{d-1} \frac{2\text{Re}(\bra{\lambda_{i}}\partial_{\mu}\hat{\rho}\ket{\lambda_{j}} \bra{\lambda_{j}} \partial_{\tau}\hat{\rho} \ket{\lambda_{i}})}{\lambda_{i}+\lambda_{j}}.
\end{equation}
The QFI for such coherent mixtures of frequency cat-like states ($\lambda_{i}=1/2$) can be cast as:
\begin{multline}
{\cal{F}}_{\tau\mu}=2\text{Re}(\bra{\psi_{\Delta}} \partial_{\tau} \hat{\rho} \ket{\psi_{\Delta}}\bra{\psi_{\Delta}} \partial_{\mu} \hat{\rho} \ket{\psi_{\Delta}}\\+\bra{\psi_{\Delta}} \partial_{\tau} \hat{\rho} \ket{\psi_{-\Delta}}\bra{-\psi_{\Delta}} \partial_{\mu} \hat{\rho} \ket{\psi_{\Delta}}\\
+\bra{\psi_{-\Delta}} \partial_{\tau} \hat{\rho} \ket{\psi_{\Delta}}\bra{\psi_{\Delta}} \partial_{\mu} \hat{\rho} \ket{\psi_{-\Delta}} \\+\bra{\psi_{-\Delta}} \partial_{\tau} \rho \ket{\psi_{-\Delta}}\bra{\psi_{-\Delta}} \partial_{\mu} \hat{\rho} \ket{\psi_{-\Delta}}).
\end{multline}
Where, for instance,
\begin{multline}
\partial_{\tau} \hat{\rho}= \frac{1}{2}(\ket{\partial_{\tau}\psi_{\Delta}}\bra{\ket{\psi_{\Delta}}}+\ket{\psi_{\Delta}}\bra{\partial_{\tau}\psi_{\Delta}}\\+\ket{\partial_{\tau}\psi_{-\Delta}}\bra{\ket{\psi_{-\Delta}}}+\ket{\psi_{-\Delta}}\bra{\partial_{\tau}\psi_{-\Delta}}).
\end{multline}
We start evaluating the diagonal terms. We obtain the sum of three terms,
\begin{multline}
{\cal{F}}_{\tau\tau}(\tau)=4 (\int d\omega_{-} \abs{f_{-}^{\Delta}(\omega_{-})}^{2} \omega_{-} \text{sin}(2\omega_{-}\tau))^{2}\\+4 (\int d\omega_{-} \abs{f_{-}^{-\Delta}(\omega_{-})}^{2} \omega_{-} \text{sin}(2\omega_{-}\tau))^{2}\\
+4\abs{\int d\omega_{-} f^{\Delta}(\omega_{-}) f_{-}^{-*\Delta}(\omega_{-})\omega_{-} \text{sin}(2\omega_{-}\tau)}^{2}\\+4(\abs{\int d\omega_{-} f^{* \Delta}(\omega_{-}) f_{-}^{-\Delta}(\omega_{-})\omega_{-} \text{sin}(2\omega_{-}\tau)}^{2}).
\end{multline}
The last expression can be written by using statistical average of the chronocyclic Wigner distribution of the phase matching function of both peak:
\begin{equation}\label{generalexpressionQFImixed}
{\cal{F}}_{\tau\tau}(\tau)=4 \sum_{i,j=1}^{2} (\iint d\omega_{-} dt \ \omega_{-} \text{sin}(2\omega_{-}\tau) W^{i,j}_{-}(\omega_{-},t))^{2},
\end{equation}
where $i,j=\pm \Delta$, and for instance $W^{\Delta,\Delta}_{-}(\omega,t)= \int d\omega_{-} e^{2i\omega' t} f^{\Delta}_{-}(\omega-\omega') f^{*\Delta}_{-}(\omega+\omega')$ and the cross Wigner distribution: $W^{\Delta,-\Delta}_{-}(\omega,t)= \int d\omega' e^{2i\omega' t} f^{\Delta}_{-}(\omega-\omega') f^{*-\Delta}_{-}(\omega+\omega')$. This formula could be generalized for others terms, such as the grid mixed state. Then, we recover Eq.~(\ref{QFImixedfreqcat}) after calculation. Note that the frequency QFI ${\cal{F}}_{\mu\mu}(\mu)$ will have a similar expression as Eq.~(\ref{generalexpressionQFImixed}).

\section{Example of calculation of variances for  grid states}\label{calculationvariance}
In this section, we show the expression of the QFI for grid states modified by a chirp (see Eq.~(\ref{multiparamatrix})). The covariance matrix of such a state can be written as:
\begin{multline}
\langle \omega_{-} t_{-} \rangle = \iint d\omega_{-} d\tau \ \omega_{-} \tau  W_{-}(\omega_{-},\tau)\\
={\cal{N}} \sum_{n,m} \iint d\omega_{-} d\tau \  \omega_{-} e^{-\omega_{-}^{2}/\sigma^{2}} e^{2i(n-m)\omega_{-} \overline{\tau}} e^{-\tau^{2}\sigma^{2}}  \\
\cross (\tau+(n+m)\overline{\tau}+\frac{\omega_{-}}{C^{2}}),
\end{multline}
where ${\cal{N}}$ is the denominator of the chronocyclic Wigner distribution (see Eq.~(\ref{frequencyphasewigner})). The first term is zero, because there is a temporal integration of the product of an odd and an even function. The second term, which is imaginary, is also zero after summation on $n$ and $m$. Only the third term is different from zero and by using the Gaussian integral 
Eq.~(\ref{Gaussianintegral2}), we obtain:
\begin{multline}
{\cal{N}} \sum_{n,m} R^{n+m} \iint d\omega_{-} d\tau e^{-\omega_{-}^{2}/\sigma^{2}} e^{2i(n-m)\omega_{-} \overline{\tau}} e^{-\tau^{2}\sigma^{2}} \frac{\omega_{-}^{2}}{C^{2}}\\
={\cal{N}} \sum_{n,m} R^{n+m} \frac{\sigma^{2}}{2C^{2}}(1-(n-m)^{2}\overline{\tau}^{2}\sigma^{2}/2)e^{-(n-m)^{2}\overline{\tau}^{2}\sigma^{2}},
\end{multline}
which gives indeed Eq.~(\ref{multiparamatrix}). This example of calculation can be applied for the other expression of the QFI for the time-frequency grid state.

\begin{figure*}
\begin{center}
 \includegraphics[scale=0.18]{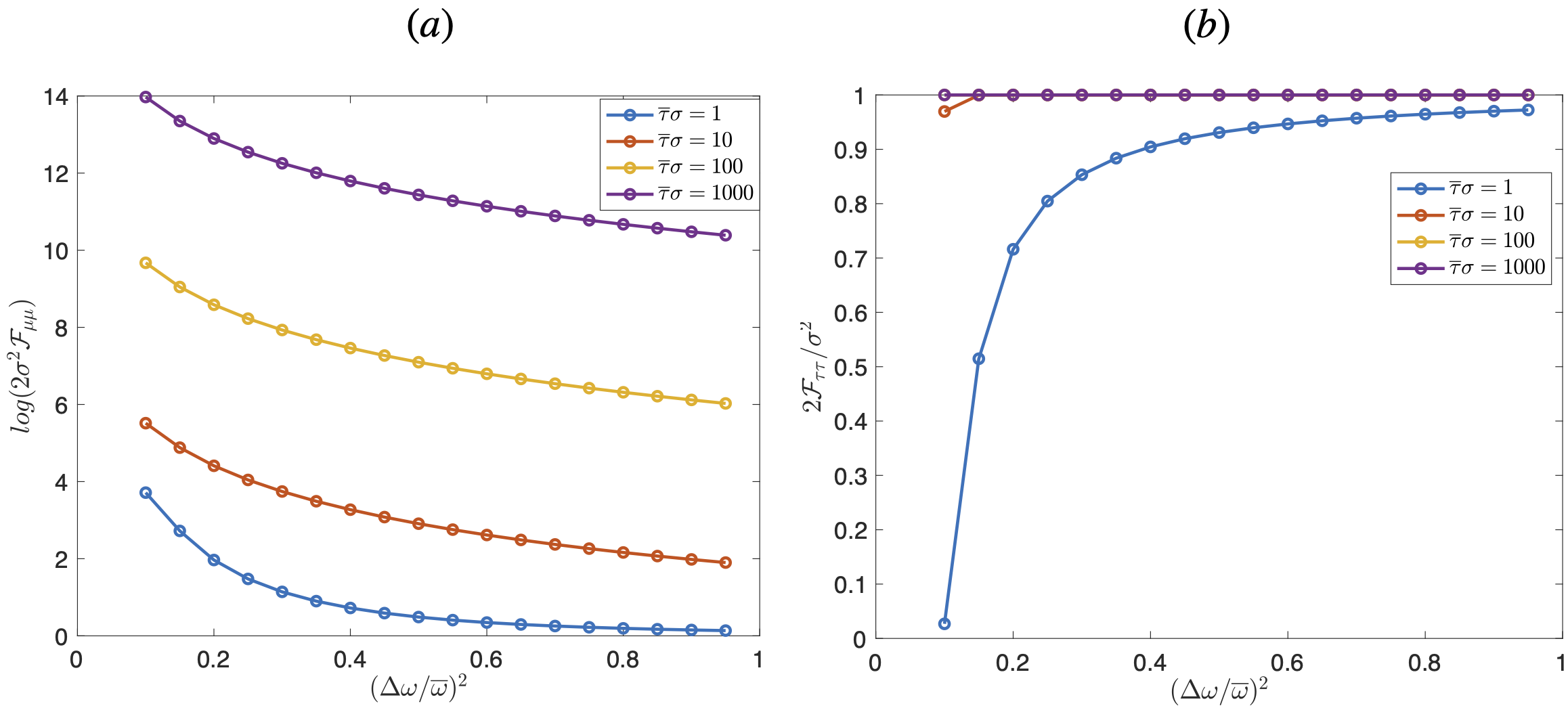}
 \caption{\label{temporalestigauss} (a) Frequency  $2{\cal{F}}_{\mu\mu}\sigma^{2}$ (in log scale) and temporal (b) Quantum Fisher information $2{\cal{F}}_{\tau\tau}/\sigma^{2}$ with respect to $\Delta\omega/\overline{\omega}$ for different value of the product $\sigma\overline{\tau}$.}
 \end{center}
\end{figure*}

\section{Saturation of the Cramer-Rao bound}\label{SaturationCramerRao}

\subsection{Variances and saturation of Cramer-Rao}
The measure for the confidence interval in the determination of $\alpha$ is given by :
\begin{equation}
\text{cov}(\alpha)=\int \prod_{k}dx_{k}p(x_{k}|\alpha)(\tilde{\alpha}-\alpha)^{2},
\end{equation}
which is bounded by the Fisher information as: $\text{cov}(\alpha)\geq 1/ NF_{\alpha}(\alpha)$.  
We will look for the optimal estimation strategy using a maximum-likelihood estimator. The likelihood estimator is a multinomial distribution $L(N_{0},N_{1},N_{2})= P_{0}^{N_{0}}P_{1}^{N_{1}}P_{2}^{N_{2}}$ and  which is extremized as:
\begin{equation}
0=\frac{N_{0} \partial_{\tau} P_{0}}{P_{0}}+\frac{N_{1} \partial_{\tau} P_{1}}{P_{1}}+\frac{N_{2} \partial_{\tau} P_{2}}{P_{2}}.
\end{equation}
This extremization  leads to an analytic expression of the optimal estimator $\tilde{\alpha}=(\tilde{\tau},\tilde{\mu})$ given by $N_{1}P_{2}(\tilde{\tau})=N_{2}P_{1}(\tilde{\tau})$. By using Eq.~(\ref{listprobability}), we find:
\begin{equation}\label{maxlikelihood}
W_-(\tilde{\tau},\tilde{\mu})=\frac{N_{2}-N_{1}\frac{1+3\gamma}{1-\gamma}}{N_{1}+N_{2}},
\end{equation}
where $N_{1}$ and  $N_{2}$ are the number of single detections and coincidence counts, respectively. This procedure  allows one to obtain an analytic expression of the optimal estimator $\tilde{\alpha}=(\tilde{\tau},\tilde{\mu})$. Indeed,  in the limit of large $N$, the probability can be written as a normal distribution, the maximum likelihood estimator is the optimal one since it saturates the Cramer Rao bound, $\Delta^{2} \alpha=1/NF_{\alpha}(\tilde{\alpha})$. Notice that it also permits  to find the photodetector losses $\gamma$ \cite{lyons_attosecond-resolution_2018}, which depends on $N_{1}$ and $N_{2}$.

\begin{figure*}
\begin{center}
 \includegraphics[scale=0.18]{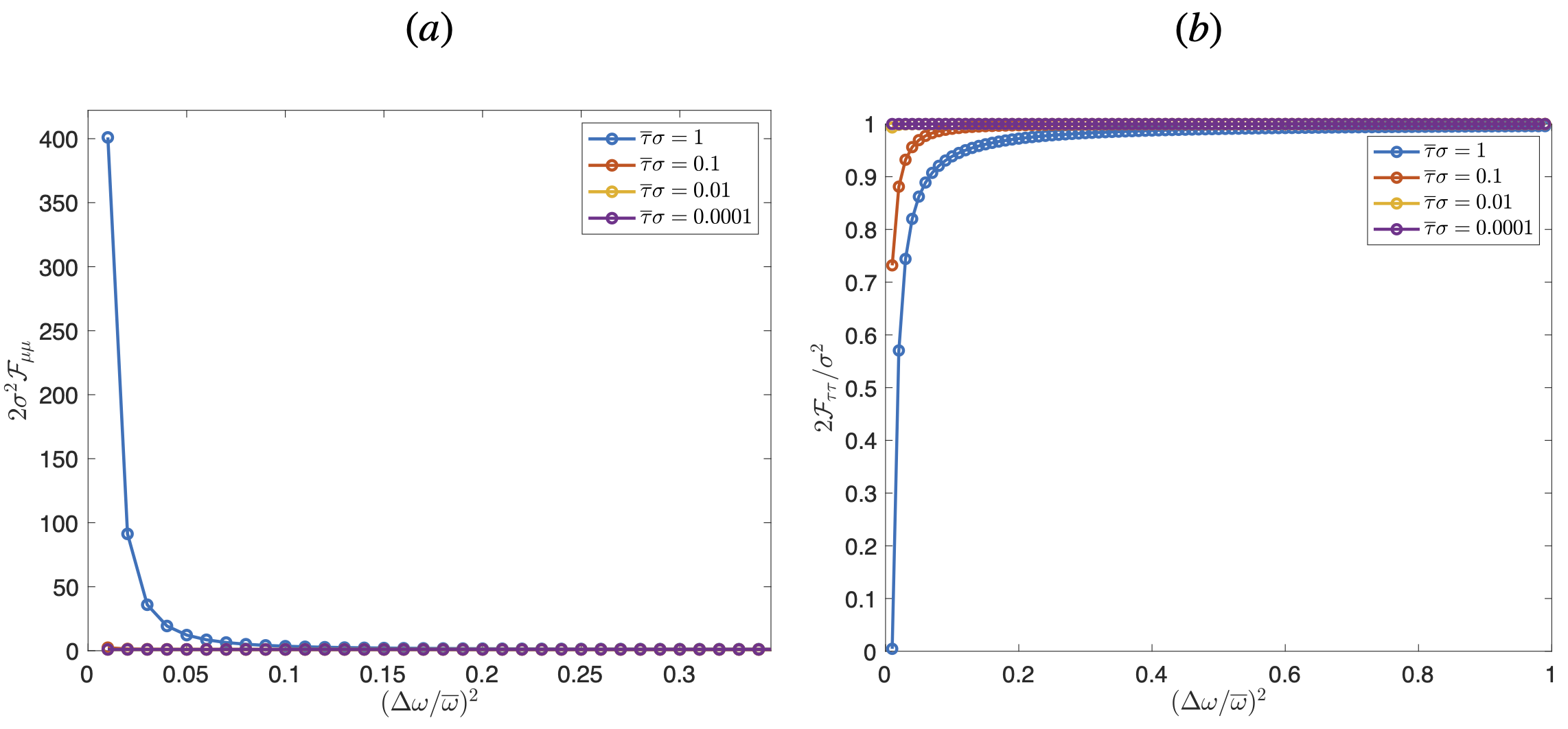}
 \caption{\label{temporalestigauss1} (a) Frequency $2{\cal{F}}_{\mu\mu}\sigma^{2}$ (in log scale) and temporal (b) Quantum Fisher information $2{\cal{F}}_{\tau\tau}/\sigma^{2}$ with respect to $\Delta\omega/\overline{\omega}$ for different value of the product $\sigma\overline{\tau}$.}
 \end{center}
\end{figure*}

We discuss now the temporal estimation, {\it{i.e.}} we set the frequency shift to zero $W_-(\tilde{\tau},0)$.  For the case where the phase-matching function is Gaussian (see Eq.~(\ref{simpleGaussian})), an analytic expression of $\tilde{\tau}$ with respect to $N_{1,2}$ and $\gamma$ can be found in \cite{lyons_attosecond-resolution_2018}.  The resolution of Eq.~(\ref{maxlikelihood}) can be found only numerically when the biphoton state is a micro-comb. The estimator is said to be efficient when it saturates the Cramer-Rao bound for a large number of repetitions of the experiment.
For large photon losses, namely when $N_{2}<N_{1}\frac{1+3\gamma}{1-\gamma}$ the right term of Eq.~(\ref{maxlikelihood}) becomes negative, the maximum likelihood technique does not allow to find $\tilde{\tau}$ for certain biphoton states which have positive Wigner distribution. In that case, other estimation techniques such as the mean square method can be applied.

In the HOM experiment, one measures the coincidence probability to obtain an estimation of  $\tau$ and $\mu$. The time or frequency precision of the non-resolved HOM experiment is given by,
\begin{equation}\label{timedelay}
\Delta \alpha= \frac{\sqrt{\langle \Delta P_{2}(\alpha,0) \rangle }}{ \abs{\frac{\partial \langle P_{2}(\alpha,0) \rangle}{\partial \alpha }}},
\end{equation}
where the variance of the non-resolved time coincidence measurement is (see Ref. \cite{chen_hong-ou-mandel_2019}) $\Delta P_{2}(\alpha,0)= P_{2}(\alpha,0)(1-P_{2}(\alpha,0))$ where again $\alpha=\tau,\mu$ and the denominator depends on the sensitivity of the mean value of the operators to the variables $\tau$ or $\mu$. Finally, for zero losses the FI is under the form,
\begin{equation}
F_{\tau\tau}(\tau)=\frac{1}{2} \frac{(\partial_{\tau} W_{-}(0,\tau))^{2}}{1-W_{-}^{2}(0,\tau)},
\end{equation}
which indeed converges towards the estimator Eq.~(\ref{timedelay}).

\section{Quantum Fisher information of Gaussian and Dirac grid states}\label{appendixgkp}

\subsection{Quantum Fisher information of the Gaussian grid state}
Let us now consider a state whose  phase matching function can be cast  as the product of an envelop of width $\sigma$ and the sum of Gaussian functions of width $\Delta\omega$,
\begin{equation}\label{gaussiansamplitude}
f_-(\omega_-)=Ae^{-\omega_{-}^{2}/2\sigma^{2}} \sum_{n=-\infty}^{\infty} e^{-(\omega_- -n\overline{\omega})^{2}/2\Delta\omega^{2}}.
\end{equation}
Here $A$ is a normalisation factor which can be expressed as $1=A^{2} \sum_{n,m} \sqrt{\frac{\pi}{1/\sigma^{2}+1/\Delta\omega^{2}}} \text{exp}(-(n-m)^{2} \overline{\omega}^{2}/4\Delta\omega^{2})$. Then, the phase-matching function is even and the QFI converges to the FI, for again zero photodetectors losses and $\tau,\mu\rightarrow 0$.
 By using the Poisson's summation formula (valid since $\abs{f_{-}(\omega_{-})}\leq C/(1+\abs{\omega_{-}}^{\alpha}))$, we can sum over $\mathds{Z}$ and the phase-matching function can be rewritten as,
\begin{equation}
f_-(\omega_-)=\frac{A}{\overline{\omega}}e^{-\omega_{-}^{2}/2\sigma^{2}}  \sum_{n\in\mathds{Z}} e^{-\frac{n^{2}}{2}(\frac{\Delta \omega}{\overline{\omega}})^{2}} e^{2in\pi \omega/\overline{\omega}}.
\end{equation}
We directly see the resemblance with the Fabry-Perot cavity model of Eq.~\eqref{phasefunction}, where $R^{n}=\text{exp}(n\text{ln}(R))\leftrightarrow e^{-\frac{n^{2}}{2}(\frac{\Delta \omega}{\overline{\omega}})^{2}}$. We can then express the Fourier transform of the phase-matching function as,
\begin{equation}
\tilde{f}_{-}(t_{-})=\frac{A}{\overline{\omega}}\sum_{n\in\mathds{Z}} e^{-\frac{n^{2}}{2}(\frac{\Delta \omega}{\overline{\omega}})^{2}}e^{-(t_{-}+\frac{2\pi n}{\overline{\omega}})^{2}\sigma^{2}/2}.
\end{equation}
We will define $\overline{\tau}=2\pi/\overline{\omega}$. In the case where $\Delta \omega \rightarrow 0$ and $\sigma\rightarrow \infty $, then the function becomes a Dirac Comb of period $\overline{\omega}$, presented in the next section. The Chronocyclic Wigner distribution can be cast as follows:
\begin{align}
W_{-}(\omega,\tau)= {\cal{N}}_{G} \sum_{n,m\in \mathds{Z}^{2}} e^{-\frac{n^{2}}{2}(\frac{\Delta \omega}{\overline{\omega}})^{2}} e^{-\frac{m^{2}}{2}(\frac{\Delta \omega}{\overline{\omega}})^{2}}\\
\cross e^{2i\pi(n-m)(\omega-\omega_{0})/\overline{\omega}} e^{-(\tau+\frac{2\pi (n+m)}{\overline{\omega}})^{2}\sigma^{2}/2},\\
{\cal{N}}_{G}= \sum_{n,m\in \mathds{Z}^{2}} e^{-\frac{n^{2}}{2}(\frac{\Delta \omega}{\overline{\omega}})^{2}} e^{-\frac{m^{2}}{2}(\frac{\Delta \omega}{\overline{\omega}})^{2}}e^{-(\frac{2\pi (n+m)}{\overline{\omega}})^{2}\sigma^{2}/2}.
\end{align}
Hence, the Quantum Fisher information has the same expression as in Eq.~(\ref{fishercomb}), but with the moments are,
\begin{align}
\langle (n\pm m)^{\alpha} \rangle = {\cal{N}}_{G}\sum_{n,m\in \mathds{Z}^{2}} (n\pm m)^{\alpha} e^{-\frac{n^{2}}{2}(\frac{\Delta \omega}{\overline{\omega}})^{2}} \\
\times e^{-\frac{m^{2}}{2}(\frac{\Delta \omega}{\overline{\omega}})^{2}} e^{-(m-n)^{2}\tau^{2}\sigma^{2}} .
\end{align}
We represent the temporal and the frequency QFI in Fig.~\ref{temporalestigauss}(a),(b) in Fig.~\ref{temporalestigauss1}(a),(b) as function of the ratio $\Delta\omega/\overline{\omega}$ for different values of the product $\overline{\tau}\sigma$. The QFI increases as $\Delta\omega/\overline{\omega}$ is closer to 0, namely that when each peak is narrower. As $e^{-\frac{n^{2}}{2}(\frac{\Delta \omega}{\overline{\omega}})^{2}}\sim1$ the state is no longer periodic.

\subsection{Dirac Comb model}\label{Diracmodel}
To understand the succession of positive and negative peaks in the chronocyclic Wigner distribution (see Fig. \ref{wigner}), we remind the Dirac comb case $f_{-}(\omega_{-})=\sum_{n\in\mathds{Z}} \delta(\omega_{-}-n\overline{\omega})$ from Ref.\cite{testorf_fractional_1996}, which present the chronocyclic Wigner distribution of a classical field with such a grid-structured spectrum.
The chronocyclic Wigner distribution of the Dirac comb is,
\begin{equation}
W_{-}(\omega,\tau)=\sum_{n,m\in\mathds{Z}^{2}} e^{2i\pi \omega /\overline{\omega}(n-m) }\delta(\tau-\frac{(n+m)}{2\overline{\omega}}).
\end{equation}
We perform the change of variable: $n'=n+m$,
\begin{equation}
W_{-}(\omega,\tau)=\sum_{n\in\mathds{Z}}\delta(\omega-n\overline{\omega}/2)\sum_{n'\in\mathds{Z}} e^{2i\pi \omega n'/\overline{\omega}} \delta(\tau-\frac{n'}{2\overline{\omega}})
\end{equation}
and we finally obtain
\begin{equation}
W_{-}(\omega,\tau)=\sum_{n,n'\in\mathds{Z}^{2}} (-1)^{nn'}\delta(\omega-n\overline{\omega}/2)\delta(\tau-\frac{n'}{2\overline{\omega}}).
\end{equation}
From this expression, the succession of positive and negative peaks is clearer. For even values of $n,n'$, the chronocyclic Wigner distribution is positive and the negative peaks do not correspond to the peaks of the marginals of the distribution.

\section{Gaussian integrals}
We remind the following Gaussian integrals, which are used all along this paper.

\begin{equation}
\int_{\mathds{R}} e^{-\alpha\omega^{2}}e^{\beta\omega}d\omega=\sqrt{\frac{\pi}{\alpha}} e^{\beta^{2}/4\alpha}
\end{equation}

\begin{equation}
\int_{\mathds{R}} \omega e^{-\omega^{2}\alpha} e^{\omega \beta}=\frac{\beta}{2\alpha}  \sqrt{\frac{\pi}{\alpha}} e^{\beta^{2}/4\alpha}
\end{equation}

\begin{equation}\label{Gaussianintegral2}
\int_{\mathds{R}}  \omega^{2} e^{-\omega^{2}\alpha} e^{\omega \beta}=\frac{1}{2\alpha} \sqrt{\frac{\pi}{\alpha}} (1+\frac{\beta^{2}}{2\alpha}) e^{\beta^{2}/4\alpha} 
\end{equation}

\bibliography{bibliometro}

\end{document}